\documentclass[11pt,article,aps,
float, 
nofootinbib]{revtex4}
\usepackage{tikz,color,cancel,acronym}
\usepackage[colorlinks,linkcolor=blue,citecolor=blue,urlcolor=blue ]{hyperref}
\usepackage{amsmath}
\usepackage{hyperref}
\usepackage{multirow}
\allowdisplaybreaks[4] 

\usepackage{graphicx}
\usepackage{changes}
\usepackage[capitalize]{cleveref}
\newcommand{\beq}{\begin{equation}}
\newcommand{\beqa}{\begin{eqnarray}}
\newcommand{\eeq}{\end{equation}}
\newcommand{\eeqa}{\end{eqnarray}}

\newcommand{\TRC}{MOE Key Laboratory of TianQin Mission, TianQin Research Center for Gravitational Physics $\&$ School of Physics and Astronomy, Frontiers Science Center for TianQin, Gravitational Wave Research Center of CNSA, Sun Yat-sen University (Zhuhai Campus), Zhuhai 519082, China}

\begin{document}

\title{Detecting the stochastic gravitational wave background with the TianQin detector}

\author{Jun Cheng }

\author{En-Kun Li}
\thanks{Corresponding author: \href{mailto:lienk@mail.sysu.edu.cn}{lienk@mail.sysu.edu.cn}}

\author{Yi-Ming Hu }
\thanks{Corresponding author: \href{mailto:huyiming@sysu.edu.cn}{huyiming@sysu.edu.cn}}

\author{Zheng-Cheng Liang }

\author{Jian-dong Zhang}

\author{Jianwei Mei }

\affiliation{\TRC}

\begin{abstract}
    The detection of stochastic gravitational wave background (SGWB) is among the leading scientific goals of the space-borne gravitational wave observatory, which would have significant impact on astrophysics and fundamental physics.
    In this work, we developed a data analysis software, \texttt{TQSGWB}, which can extract isotropic SGWB using the Bayes analysis method based on the TianQin detector.
    We find that for the noise cross spectrum, there are imaginary components and they play an important role in breaking the degeneracy of the position noise in the common laser link.
    When the imaginary corrections are considered, the credible regions of the position noise parameters are reduced by two orders of magnitude.
    We demonstrate that the parameters of various signals and instrumental noise could be estimated directly in the absence of a Galactic confusion foreground through Markov chain Monte Carlo sampling.
    With only a three-month observation, we find that TianQin could be able to confidently detect SGWBs with energy density as low as $\Omega_{\rm PL} = 1.3 \times 10^{-12}$, $\Omega_{\rm Flat} = 6.0 \times 10^{-12}$, and $\Omega_{\rm SP} = 9.0 \times 10^{-12}$, for power-law, flat, and single-peak models respectively.
\end{abstract}

\maketitle

\acrodef{GW}{gravitational wave}
\acrodef{GCB}{Galactic ultra compact binaries}
\acrodef{SGWB}{Stochastic gravitational wave background}
\acrodef{SNR}{signal-to-noise ratio}
\acrodef{DWD}{double white dwarf}
\acrodef{MBHB}{massive black hole binary}
\acrodefplural{MBHB}[MBHBs]{massive black hole binaries}
\acrodef{SBBH}{stellar-mass black hole binary}
\acrodefplural{SBBH}[SBBHs]{stellar-mass black hole binaries}
\acrodef{EMRI}{extreme mass ratio inspiral}
\acrodef{PTA}{pulsar timing array}
\acrodef{CE}{cosmic explorer}
\acrodef{ET}{Einstein telescope}
\acrodef{LISA}{laser interferometer space antenna}
\acrodef{O1}{first observing run}
\acrodef{O3}{third observing run}
\acrodef{LVK}{the LIGO Scientific Collaboration, the Virgo Collaboration and the KAGRA Collaboration}
\acrodef{PCA}{principal component analysis}
\acrodef{CBC}{compact binary coalescences}
\acrodef{TDI}{time delay interferometry}
\acrodef{PSD}{power spectral density}
\acrodef{SNR}{signal-to-noise ratio}
\acrodef{PDF}{probability distribution function}
\acrodef{MCMC}{Markov Chain Monte Carlo}
\acrodef{NS}{nested sampling}

\section{Introduction}

\ac{SGWB} \cite{Allen:1997ad,Romano:2016dpx,Christensen:2018iqi}
can be produced by the incoherent superposition of large numbers of independent and unresolved gravitational wave sources.
According to the different physical mechanisms and historical epochs that contributed to the SGWB,
it could be classified into two major types:

\paragraph{Astrophysical origin.}
This class includes a Galactic foreground contributed from Galactic  \acp{DWD} \cite{Huang:2020rjf},
and extragalactic backgrounds produced by the inspiral and merger of compact binaries,
such as \acp{MBHB} \cite{Wang:2019ryf},
\acp{SBBH}  \cite{Liu:2020eko,Liu:2021yoy},
and \acp{EMRI} \cite{Fan:2020zhy,Zi:2021pdp}.

The Galactic foreground is comparable with detector noise, and thus is the most distinctive.
The Galactic origin also made it anisotropic.
On the other hand, the extragalactic background is generally expected to be spatially homogeneous, Gaussian, stationary, unpolarized, and isotropic.
The detection of such SGWB can provide important astrophysical information about the source populations,
like the mass distribution of the binary, the merger rate evolution, and even the formation mechanisms \cite{Mazumder:2014fja,Callister:2016ewt,Maselli:2016ekw}.

\paragraph{Cosmological origin.}
It has been suggested that various processes in the early Universe\cite{Maggiore:2000gv,Caprini:2018mtu},
including inflation \cite{Guth:1982ec,Bartolo:2016ami},
first-order phase transition \cite{Hogan:1983ixn,Caprini:2015zlo,Caprini:2019egz},
and networks of topological defects (e.g., cosmic strings) \cite{Kibble:1976sj,Auclair:2019wcv} can contribute to the SGWB.

The cosmological SGWB thus encodes precious information about the early Universe, and the detection of it has the potential to significantly improve our understanding of the processes that shaped the early Universe and particle physics beyond the standard model \cite{ValbusaDallArmi:2020ifo}.
For example, one can decipher the energy scale and slope of inflation potential, as well as the initial quantum states during inflation by observing the relic gravitational wave (RGW) \cite{Wang:2018arb}.

The search for \ac{SGWB} with current experiments, like
Advanced LIGO \cite{LIGOScientific:2014qfs},
Advanced Virgo \cite{VIRGO:2014yos},
and \acp{PTA} \cite{article,Detweiler:1979wn,osti_6335283} are carried out actively.
Meanwhile, relevant investigations with future experiments, like
the \ac{ET} \cite{Punturo:2010zz},
the \ac{CE} \cite{LIGOScientific:2016wof,Reitze:2019iox},
as well as space-borne missions such as \ac{LISA}  \cite{LISA:2017pwj}
and TianQin \cite{Luo:2015ght} are also performed.
With data accumulated during Advanced LIGO and Advanced Virgo's \ac{O1} till the \ac{O3}, no observational evidence for \ac{SGWB} has been reported offically from \ac{LVK}, which places an  upper limit on the dimensionless energy density $\Omega_{\rm gw} \le 5.8 \times 10^{-9}$ for a frequency-independent (flat) \ac{SGWB} \cite{KAGRA:2021kbb}.
Recently, the NANOGrav Collaboration reported an interesting common-spectrum process that can not be easily explained by noise, yet no strong support of quadrupolar spatial correlations can be concluded \cite{NANOGrav:2020bcs}.
Such observations have been later confirmed by PPTA \cite{Goncharov:2021oub} and EPTA \cite{Chen:2021rqp}.
Many hypotheses have been proposed to explain the observed common-spectrum process, including cosmic strings, dark phase transitions, or scalar transverse polarization mode \cite{Ellis:2020ena,Addazi:2020zcj,Ratzinger:2020koh,Blasi:2020mfx,Buchmuller:2020lbh,Samanta:2020cdk,Nakai:2020oit,Neronov:2020qrl,Chen:2021wdo}.
But more observation is needed to draw definitive conclusions.

Since the statistical property of the \ac{SGWB} is indistinguishable from the instrumental noise, it is challenging to identify \ac{SGWB} with one interferometer, unless the strength is comparable or even larger than the instrumental noise, as in the case of the Galactic foreground.
In principle, one can use the cross-correlation method to identify the existence of \ac{SGWB} with multiple detectors, using the fact that \ac{SGWB} recorded by multiple detectors are correlated, while noise in different detectors are statistically independent \cite{Hellings:1983fr,Christensen:1992wi,Flanagan:1993ix,Allen:1997ad,Seto:2020zxw}.
For the special case of a triangular-shaped \ac{GW} detector, one can construct a signal-insensitive data stream, also known as the \emph{null channel}, which has been suggested that can play the role of a noise monitoring channel, and can be further used to search for \ac{SGWB} \cite{Estabrook:2000ef,Hogan:2001jn}.
The validity of such method had been established through the analysis on the mock LISA data challenge \cite{Adams:2010vc,Adams:2013qma}.

Many interesting works have emerged for more realistic and robust studies of \ac{SGWB}.
For example, \cite{Pieroni:2020rob} proposed a search method based on \ac{PCA}, so that no \emph{a priori} assumption of \ac{SGWB} spectral shape is needed.
Other spectrum-agnostic methods have also been developed \cite{Karnesis:2019mph,Caprini:2019pxz,Flauger:2020qyi}.
Methods like adaptive \ac{MCMC} have been proposed to identify multicomponent backgrounds \cite{Boileau:2020rpg,Boileau:2021sni}.
And advances have been made to identify the anisotropy of \ac{SGWB} \cite{Seto:2004np,Alonso:2020mva,LISACosmologyWorkingGroup:2022kbp,KAGRA:2021mth}

In this paper we focus on the mHz frequency band and developed a search method for \ac{SGWB} with space-borne \ac{GW} missions like TianQin \cite{Luo:2015ght,Hu:2017yoc,TianQin:2020hid}, which we call \texttt{TQSGWB}.
We study the detection capability of TianQin to \ac{SGWB} in the absence of Galactic confusion foreground.
We find that the imaginary part of the noise cross spectrum, which was mostly overlooked in the field, can break the degeneracy of position noise in common laser link.
We demonstrated the detection capability of TianQin for three representative \ac{SGWB} spectra, as well as its detection limits.

This paper is organized as follows. In Sec. \ref{sec:SGWB} we introduce the fundamentals of the \ac{SGWB} and the three representative spectra.
In Sec. \ref{sec:Detector}, we describe in detail the noise model of the TianQin detector and its response function.
In Sec. \ref{sec:SEARCH METHOD}, we show how one could simulate the data stream by using the power spectral density of SGWB and noise,
and also briefly describe the Bayesian analysis technique, as well as our main results.
We finally draw our conclusions and discuss possible extensions of this work in Sec. \ref{sec:SUMMARY AND DISCUSSION}.

\section{Fundamentals of Stochastic gravitational-wave background }
\label{sec:SGWB}

In the transverse-traceless gauge, the metric perturbations $h_{ij}(t,\vec{x})$ corresponding to SGWB can be written as a
superposition of plane waves having frequency $f$ and propagating in the direction of $\hat{k}$,
\beq
\label{expa}
h_{ij}(t,\vec{x}) =
\int_{-\infty}^{\infty} df
\int_{S^{2}} d{\hat{k}} \, h_{ij}(f,\hat{k})\, e^{ {\rm i}2\pi f(t-\hat{k}\cdot \vec{x}/c)},
\eeq
where $h_{ij}(f,\hat{k})=\sum_{P} \,h_{P}(f,\hat{k})\,e_{ij}^P(\hat{k})$, $P=\{+,\times\}$ denotes polarization states,
and $e_{ij}^P$ are the polarization tensors.

Considering a Gaussian, stationary, unpolarized, spatially homogenous, and isotropic \ac{SGWB},
the statistical properties of the corresponding Fourier components 
satisfy \cite{Romano:2016dpx,Allen:1997ad}
\begin{align}
    \left\langle h_{P}(f,\hat k) \right\rangle =& \, 0,
    \\
    \left\langle h_{P}(f,\hat k) h_{P'}^*(f',\hat k') \right\rangle
    =& \frac{1}{16\pi} S_h(f) \delta(f-f') \delta_{PP'} \delta^2(\hat k,\hat k'), \label{eq:sh}
\end{align}
where $\langle ... \rangle$ denotes the ensemble average, and $S_h(f)$ is the one-sided power spectral density (PSD),
which is related to the fracional energy density spectrum $\Omega_{\rm gw}(f)$ of SGWB as
\beq
S_h(f)=\frac{3H_0^2}{2\pi^3} \frac{\Omega_{\rm gw}(f)}{f^{3}}
\text{ and }
\Omega_{\rm gw} (f) = \frac{1}{\rho_c} \frac{d\rho_{\rm gw}}{d\ln f}.
\eeq
Here, $\rho_{\rm gw}$ is the GWs energy density, $\rho_c = 3H_0^2/(8\pi G)$ represents the critical density,
and $H_0$ is the Hubble constant, for which we assume a fiducial value of $H_0=67 \,{\rm km\, s^{-1} Mpc^{-1}}$ throughout this work \cite{Planck:2015fie}.

We consider three scenarios as follows: \ac{SGWB} from the \acp{CBC} (power-law), a scale-invariant inflationary \ac{SGWB} (flat), and a single-peak spectrum (Gaussian-bump).

\begin{itemize}
\item{Power-law spectrum}

  \acp{CBC} are among the most promising \ac{GW} sources for the mHz frequency band  \cite{Liu:2020eko,Sesana:2016ljz,Hu:2017yoc}.
    Based on the numerous observations from current ground-based GW detectors \cite{LIGOScientific:2018mvr,LIGOScientific:2020stg,LIGOScientific:2020zkf,LIGOScientific:2020iuh,LIGOScientific:2020aai}, preliminary studies reveals that \ac{CBC} can form detectable \ac{SGWB} for TianQin \cite{Liang:2021bde}.
The analytic model describing the CBCs background depends on redshift and merger rates \cite{LIGOScientific:2017zlf,Regimbau:2011rp},
whereas the energy of the inspirals can be described by a power-law (PL) spectrum \cite{Regimbau:2011rp,Farmer:2003pa,Moore:2014lga}
\beq \label{eq:PL_signal}
\Omega_{\rm CBCs}(f) = \Omega_{\rm PL} \left( \frac{f}{f_\mathrm{ref}} \right)^{2/3},
\eeq
where $\Omega_{\rm PL}$ denotes the amplitude level at reference frequency $f_\mathrm{ref}$.
    We inject a power-law \ac{SGWB} with a fiducial value of amplitude $\Omega_{\rm PL} = 4.4 \times 10^{-12}$ and $f_\mathrm{ref} = 3 \,\mathrm{mHz}$ \cite{LIGOScientific:2019vic,Chen:2018rzo}.

\item{Flat spectrum}

  The slow-roll inflation can produce a cosmological-originated \ac{SGWB} through the amplification of vacuum fluctuations \cite{Grishchuk:1974ny,Grishchuk:1993ds,Starobinsky:1979ty,Maggiore:1999vm}.
The spectrum and amplitude of such signals strongly depend on the fluctuation power spectrum developed during the early inflationary period.
In the TianQin detection frequency regime, the inflationary background is expected to be scale invariant (flat)
\beq \label{eq:flat_signal}
\Omega_{\rm Inflation}(f)=\Omega_{\rm Flat}.
\eeq
And we choose $\Omega_{\rm Flat} = 1.0 \times 10^{-11} $ as the fiducial value.

\item{Gaussian-bump spectrum}

  Many physical mechanisms in the very early Universe can produce \ac{SGWB} with a spectrum peak around certain frequencies, including nonperturbative effects during postinflationary preheating,
strong first-order phase transitions during the thermal era of the Universe \cite{Kamionkowski:1993fg,Caprini:2007xq,Huber:2008hg,Hindmarsh:2013xza,Hindmarsh:2015qta,Caprini:2015zlo},
    or merging of primordial black holes (PBHs) during the early Universe \cite{Caprini:2015zlo,Caprini:2019egz}.
In the following analysis, we consider a phenomenological spectrum of a Gaussian-bump (or single-peaked) model \cite{Namba:2015gja,Thorne:2017jft},
\beq  \label{eq:SP_signal}
\Omega_{\rm Bump} = \Omega_{\rm SP} \exp\left\{ - \frac{ \big[ \log_{10} ( f /f_\mathrm{ref} ) \big]^2}{ \Delta^2 } \right\},
\eeq
and we adopt $\Omega_{\rm SP} = 1.0 \times 10^{-11}$, $\Delta = 0.2$, $f_\mathrm{ref} = 3 \ \mathrm{mHz}$ as the input parameters \cite{Caprini:2019pxz,Flauger:2020qyi}.

\end{itemize}

We remind the readers that the Galactic foreground from the \acp{DWD} can be dominating in certain frequency range.
However, some assumptions adopted in our following analysis do not hold for the Galactic foreground (as it is anisotropic and strong).
Therefore we concentrate on the background alone, and will leave the more realistic analysis of both the Galactic foreground and isotropic backgrounds in the future.

\section{Principals for SGWB detections with TianQin}\label{sec:Detector}

Throughout this work, we consider TianQin as our fiducial detector.
It has been proposed that the successful operation of TianQin can reveal great amount of exciting science, such as the origin and evolution of black holes and multi black hole systems \cite{Wang:2019ryf}, the astrophysics of compact binaries\cite{Liu:2020eko}, the surroundings and nature of black holes\cite{Shi:2019hqa,Zi:2021pdp}, the nature of the gravity \cite{Bao:2019kgt}, the expansion of the Universe \cite{Zhu:2021aat,Zhu:2021bpp}, and so on \cite{TianQin:2020hid,Liang:2021bde}.
The TianQin mission will be composed with three drag-free satellites, forming an equilateral triangle constellation orbiting the Earth with an orbital radius of about $10^5$ km, and it adopts a three months on and three months off (3+3) operational model \cite{Ye:2019txh,Zhang:2020paq}.


\subsection{Noise model}

Laser links will be established between TianQin satellites,  and the detection of \ac{GW} signal is implemented through laser interferometry.
Let us consider the laser link from satellite $i$ to satellite $j$: the time series of the recorded laser phase $\Phi_{ij}(t)$ are composed with contributions from possible \ac{GW} signals $\psi_{ij}(t)$ and noise, while the noise can be further classified into laser frequency noise $C$, position noise $n^p$, and acceleration noise $\vec{n}^a$:
\beqa
\Phi_{ij}(t) &=& C_i(t-L_{ij}/c)-C_j(t) + \psi_{ij}(t)+n^p_{ij}(t)
\nonumber \\
&-& \hat{r}_{ij}\cdot\big[\vec{n}^a_{ij}(t)-\vec{n}^a_{ji}(t-L_{ij}/c)\big].
\label{eq:Phase_ij}
\eeqa
Here $\hat{r}_{ij}$ is the unit vector from satellite $i$ pointing to satellite $j$,
$L_{ij}$ is the arm-length between satellite $i$ and $j$, $c$ is the light speed.
The GW strain is given by
\beq
\psi_{ij}(t) = \frac{h(f,t-L/c, \vec{x}_{i}) : (\hat{r}_{ij} \otimes \hat{r}_{ij} ) \mathcal{T}(f,\hat{r}_{ij} \cdot \hat{k}) }{2L},
\eeq
where $\mathcal{T}(f,\hat{r}_{ij} \cdot \hat{k})$ is the \emph{transfer function} of the interferometer to the \acp{GW}
for each arm \cite{Schilling:1997id,Cornish:2001qi,Cornish:2001bb,Cornish:2002rt}.

The laser frequency noise $C(t)$ is usually several orders of magnitude higher than other noise.
If left untreated, the laser frequency noise will dominate the data and make \ac{GW} impossible.
To suppress the laser frequency noise, the \ac{TDI} technique \cite{Tinto:1999yr,Estabrook:2000ef,Tinto:2002de,Prince:2002hp,Tinto:2020fcc} has been proposed and applied in the analyzing of space-borne \ac{GW} detections.
The principle of \ac{TDI} is to combine multiple time-shifted laser links into equivalent equal-arm interferometer to cancel the laser frequency noise.
Depending on the complexity of the considered satellite motion, the \ac{TDI} combinations for a Michelson interferometer can be classified into different generations;
\ac{TDI} generation 1.0 assumes a static configuration, \ac{TDI} generation 1.5 considered a rigid but rotating constellation, and \ac{TDI} generation 2.0 takes the constant changing of arm-length into account \cite{Bayle:2018hnm, Tinto:2020fcc}.
In this work, we concentrate on \ac{TDI} generation 1.5 for our analysis.

With the output phases from six different links, it is possible to built three virtual equal arm channels which are called channel $X$, $Y$, and $Z$.
We illustrate the Michelson type variable of $X$ channel with the \ac{TDI} generation 1.5 as \cite{Adams:2010vc}
\begin{equation}
    \begin{aligned}
        X(t) =& \big[\Phi_{12}(t-3L/c)+\Phi_{21}(t-2L/c)
        \\
        & + \Phi_{13}(t-L/c)+\Phi_{31}(t)\big]
        \\
        & - \big[\Phi_{13}(t-3L/c)+\Phi_{31}(t-2L/c)
        \\
        & + \Phi_{12}(t-L/c)+\Phi_{21}(t)\big].
    \end{aligned}
\end{equation}
The Fourier transform of the \ac{TDI} variable $X(f)$ is
\beqa\label{TDI-1.5X}
X_{a}(f) &=& 4{\rm i}\sin ue^{-2{\rm i}u}\big[(n^a_{31}+n^a_{21})\cos u-(n^a_{12}+n^a_{13})\big],
\nonumber \\
X_{p}(f) &=& 2{\rm i}\sin ue^{-2{\rm i}u}\big[ (n^p_{31}-n^p_{21})e^{{\rm i}u}+(n^p_{13}-n^p_{12}) \big].
\nonumber \\
\eeqa
Here $u = f/f_*$, and $f_* = c/(2\pi L)$ is the characteristic frequency.
The others \ac{TDI} variables $Y$ and $Z$, can be obtained by cyclic permutation of indices: $1 \rightarrow 2 \rightarrow 3 \rightarrow 1$ in Eqs. (\ref{TDI-1.5X}).

\begin{figure}[htbp]
    \includegraphics[width=0.6\linewidth,trim={0.1cm 0 1.2cm 1cm},clip]{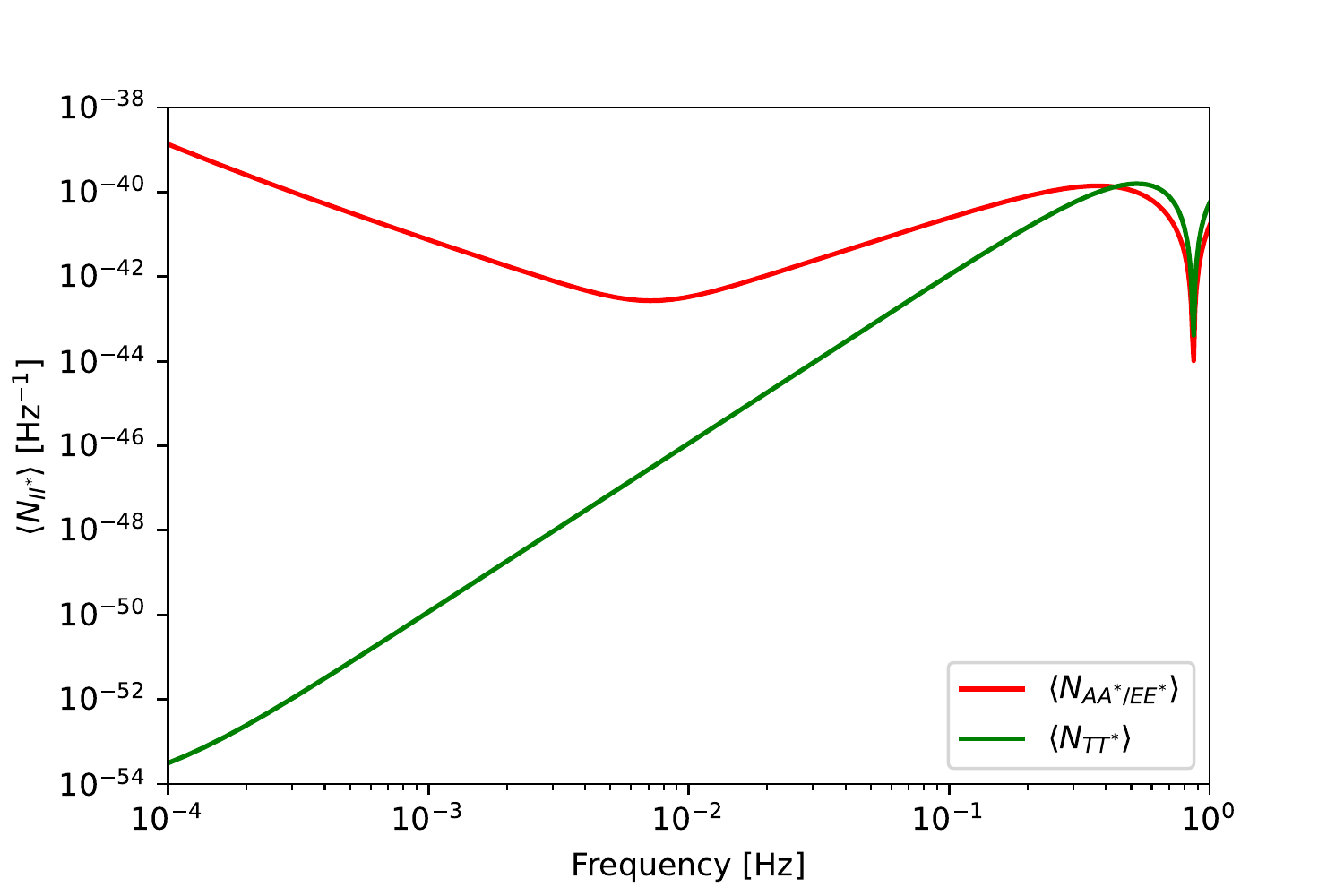}
    \caption{The noise auto-power spectral density of TianQin in the different channels.}
    \label{fig:PSD}
\end{figure}

Under the assumption that the position and acceleration noise for all the arms are identical and uncorrelated, three orthogonal TDI channels $A$, $E$, and $T$ could be constructed \cite{Prince:2002hp,Tinto:2020fcc,Vallisneri:2007xa},
\beq\label{AET}
A =\frac{Z-X}{\sqrt{2}},\, 
E =\frac{X-2Y+Z}{\sqrt{6}},\,
T =\frac{X+Y+Z}{\sqrt{3}}.
\eeq
The $A$ and $E$ channels are sensitive to \ac{GW} signals, can has similar properties to Michelson channels, but the $T$ channel is signal insensitive under the low-frequency approximation (or long-wavelength limit), and is also referred to as the \emph{null channel} or noise monitoring channel.
The noise \ac{PSD} of each channel can than be analytically expressed as \cite{Flauger:2020qyi}
\begin{equation}
    \label{AETpsd}
    \begin{aligned}
        \left\langle N_{AA^*} \right\rangle
        =& \left\langle N_{EE^*}\right\rangle
        = 8 \sin^2 u
        \bigg\{ \big[ 2+\cos u\big] S^{p}
        \\
        & + 4 \big[ 1 + \cos u + \cos^2 u\big] S^{a} \bigg\},
        \\
        \left\langle N_{TT^*}\right\rangle
        =& 32 \sin^2 u \sin^2\left( \frac{u}{2} \right)
        \bigg\{ S^{p} + 2\big[1-\cos u\big] S^{a}\bigg\}.
    \end{aligned}
\end{equation}
Here, $S_{ij}^{a}(f)=S^{a}(f) = \left\langle  \vec{n}^a_{ij}(f)\,\vec{n}^{a*}_{ij}(f)\right\rangle  $, and $ S_{ij}^{p}=S^{p}(f) = \left\langle  n^p_{ij}(f)\, n^{p*}_{ij}(f)\right\rangle  $
are nominal spectral density of acceleration noise and position noise respectively,
\begin{equation}
    \label{tqspsa}
    \begin{aligned}
        S^{p}(f) =& N^p \frac{\mathrm{m}^{2}}{\mathrm{Hz}}\left(\frac{1}{2L}\right)^{2},
        \\
        S^{a}(f) =& N^a \frac{\mathrm{m}^{2}}{\mathrm{s}^{4}\,\mathrm{Hz}}
        \left(1+ \frac{0.1 \mathrm{mHz}}{f}  \right)
        \left(\frac{1}{2\pi f}\right)^4 \left(\frac{1}{2L}\right)^{2},
    \end{aligned}
\end{equation}
where $N^p =1.0 \times 10^{-24}$ and $N^a = 1.0 \times 10^{-30}$ are the noise amplitude parameters.
These TianQin noise \acp{PSD} are shown in Fig. \ref{fig:PSD}.
With the cancellation of laser frequency noise, the acceleration noise dominates the noise at low frequencies,
while position noise dominates at high frequencies.

Upon this point, we have adopted an important assumption, that noises from different arms are identical, $S_{ij}^{a}=S^{a}$, and $ S_{ij}^{p}=S^{p}$,
naturally the cross spcetrum components $\langle N_{AE*}\rangle$ will vanish.
However, if we account for the differences between different arms, then one must account for the cross spcetral components. 
The details of all the components of the cross spcetrum matrix are shown in the Appendix~\ref{Cross-Spectra}.
It is worth emphasizing that the position noise contributions 
have imaginary components, which mainly comes from $e^{{\rm i}u}$ in the second term of Eqs. (\ref{TDI-1.5X}), for example,
\begin{equation}
    \begin{aligned}
        \left\langle N_{AE^*}^p \right\rangle
        =& \frac{2 \sin^2 u}{\sqrt{3}} \bigg\{
            \big[1 + 2\cos u \big]
            ( S_{23}^p+S_{32}^p- S_{12}^p-S_{21}^p ) \\
        +& {\color{red} {\rm i} 2 \sin u
            \left(S_{12}^p-S_{21}^p-S_{13}^p+S_{31}^p+S_{23}^p-S_{32}^p\right)}\bigg\}.
    \end{aligned}
    \label{eq:Np_AE}
\end{equation}

From Eqs.~\eqref{eq:Np_CS} and \eqref{eq:Np_AE}, one can observe that the sums of the position and acceleration noise contributions in each arm of the interferometer, e.g., $S_{13}^{a,p} + S_{31}^{a,p}$, always appear together in the autocorrelation PSDs and the real part of some cross spcetra components.
On the other hand, the differences of the position noise contributions in each arm appear only in the imaginary part of the cross spcetra,e.g., $S_{12}^p - S_{21}^p$.
This suggest that if only the real components are considered, then certain parameters are degenerate, like $S^p_{12}$ and $S^p_{21}$.
However, the consideration of the imaginary component can break such degeneracy.
Thus, one can expect that the constraints on the position noise parameters will be better than the acceleration noise parameters.

\subsection{Response function}

In general, the output of each channel in an interferometer consists of the instrumental noise and the signal,
which can be expressed in the frequency domain as
\beq
d_{I}(f) = n_{I}(f)+ h_{I}(f), \,\, I = \{A, E, T\}.
\eeq
Here, $n(f)$ is the instrumental noise, and $h(f)$ represents the interferometric response to the SGWB
\beq\label{responsesig}
h(f)=\sum_{P} \int_{S^{2}} d{\hat{k}} \, h_{P}(f,\hat{k})F^{P}(f,\hat{k})
e^{-{\rm i}2\pi f\hat k\cdot \vec x /c},\,
\eeq
with antenna pattern function $F^{P}(f,\hat{k})$.
We can express the auto- and cross spectra of stationary noise and signal as
\beqa
\langle n_{I}(f) n_{J}^*(f\rq{})\rangle
&=& \frac{1}{2}\,\delta(f-f\rq{})\,\left\langle  N_{IJ^*} \right\rangle , \\
\langle h_{I}(f) h_{J}^*(f\rq{})\rangle
&=& \frac{1}{2}\,\delta(f-f\rq{})\,{R}_{IJ^*}(f)\,S_h(f),
\eeqa
where $R_{IJ^*}(f)$ is the overlap reduction function \cite{Finn:2008vh},
\beq \label{responsefunction}
R_{IJ^*}(f) =  \frac{1}{8\pi}  \sum_{P} \int_{S^2}d{\hat k} F_I^P(f,\hat k) F_{J}^{P*}(f,\hat k) e^{ -{\rm i}2\pi f\hat{k}(\vec{x}_I-\vec{x}_J)}.
\eeq

In the equal arm-length limit, all of the overlap reduction functions $R_{AE^*}(f)$, $R_{AT^*}(f)$, and $R_{ET^*}(f)$ are zero, that is,
$A$, $E$, and $T$ channels have uncorrelated responses to an isotropic and unpolarized SGWB \cite{Adams:2010vc,Romano:2016dpx}.
As for the average response function $R_{II^*}(f)$, semianalytical expressions \cite{Larson:1999we,Larson:2002xr,Zhang:2019oet,Liang:2019pry}
or numerical methods \cite{Blaut:2012zz,Tinto:2010hz} were usually used.
Recently, a fully analytical expression of Michelson type TDI combinations has been derived \cite{Lu:2019log,Zhang:2020khm,Wang:2021jsv,Wang:2021owg,Wang:2022nea}, which provides a convenient way to calculate the response, and is adopted in this paper.

\begin{figure}[htbp]
 \includegraphics[width=0.6\textwidth]{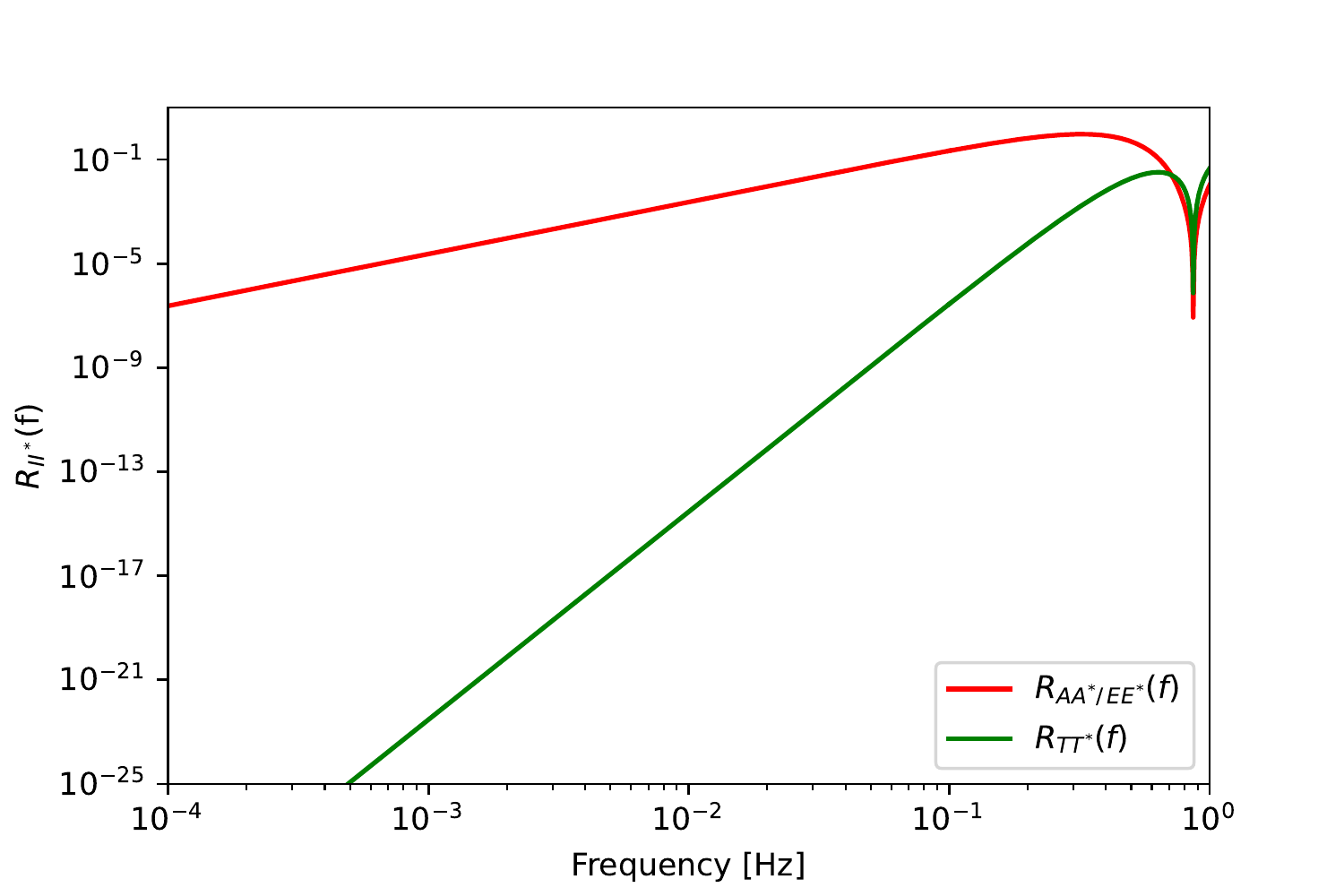}
\caption{Analytic average response function for each channel.
    Compared with the $A/E$ channel, the response of the $T$ channel to \acp{GW} is relatively negligible at low frequencies.}
\label{fig:Response}
\end{figure}

The analytic response function for each channel are shown in Fig. \ref{fig:Response}.
Compared with the $A/E$ channel, the response of the $T$ channel is negligible for frequencies below the
characteristic frequency ($f^{\rm {TQ}}_{*} \approx 0.28 $ Hz), but needs to be considered at high frequencies.

\section{Search method and results}\label{sec:SEARCH METHOD}

\subsection{Bayesian inference}
The framework of Bayesian inference is widely used in the community of astronomy, which formulates way to assess the probability distribution $p(\vec{\theta} \vert{D})$ over parameter space $\vec{\theta}$ with the observed data $D$.
The core of Bayesian inference is the Bayes' theorem,
\beq
p(\vec{\theta} \vert{D}) = \frac{{\cal L}(\vec{\theta}|D) \, p(\vec{\theta})}{p(D)},
\eeq
where $p(\vec{\theta} \vert{D})$, ${\cal L}(\vec{\theta}|D)=p(D|\vec{\theta})$,  $p(\vec{\theta})$, and $p(D)=\int  {\cal L}(\vec{\theta}|D) \, p(\vec{\theta}) d{\vec{\theta}}$ are posterior, likelihood, prior, and evidence, respectively.
The evidence plays a role of normalizing the posterior over parameter space.

Despite the relatively simple expression, it is not straightforward to map the posterior distribution over parameter space efficiently.
In practice, stochastic sampling methods like \ac{MCMC} have been widely adopted to numerically approximate the posterior distribution.
Brute force methods like grid-based search suffers from the ``curse of dimensionality", while \ac{MCMC} can efficiently sample in high-dimensional parameter space.
Unlike brute force methods, once tuned, the \ac{MCMC} methods can efficiently sample the parameter space, therefore we adopt the \ac{MCMC} to obtain the posterior distribution.
Specifically, we choose the popular implementation of the affine invariant ensemble sampler, \texttt{emcee} \cite{Foreman-Mackey:2012any} for this work.

In addition to the posterior distribution, one also face the problem of model selection, or to identify which model is better supported by the observed data.
This is usually done by calculating the {\em odds ratio} $\frac{p({\cal M}_1|D)}{p({\cal M}_0|D)}$ between the models ${\cal M}_1$ and ${\cal M}_0$.
We observe
\beq
\frac{p({\cal M}_1|D)}{p({\cal M}_0|D)} = \frac{p(D|{\cal M}_1)}{p({D|\cal M}_0)}  \frac{p({\cal M}_1)}{p({\cal M}_0)},
\eeq
which states that the odds ratio can be expressed as the product of the {\em Bayes factor} $\frac{p(D|{\cal M}_1)}{p({D|\cal M}_0)}$ and the {\em prior odds} $\frac{p({\cal M}_1)}{p({\cal M}_0)}$.
We set the prior odds to unit so that the data can dominate, and therefore we focus on the Bayes factor for the remaining of the work.
The calculation of Bayes factor is enabled through calculating of evidence under both models using the \ac{NS} algorithm \texttt{dynesty} \cite{2020MNRAS.493.3132S}

Throughout this work, we set the prior on the spectral index to be uniform, and the prior on other parameters to be logarithmically uniform ,
such as $\log_{10}\Omega_{\rm PL}\in {\cal U}[-15, -9]$, $\log_{10} S^{a}_{ij} \in {\cal U}[-43, -39]$, $\log_{10} S_{ij}^{p} \in {\cal U}[-53, -48]$.

\subsection{Likelihood}

For a stationary and Gaussian noise, the likelihood function can be expressed as \cite{Romano:2016dpx, Adams:2010vc}
\beq\label{eq:lnlikeli}
{\cal L}(\vec{\theta}|D)
=\prod_{k} \frac{1}{ \sqrt{ (2\pi)^3 |C(f_{k})| } }
\exp{\left\{-\frac{1}{2} D \left[C(f_{k})\right]^{-1} D^{\dagger}\right\}},
\eeq
where $D =[A(f_{k}), E(f_{k}), T(f_{k})]$ is the frequency-domain strain data stream from the three channels, $^{\dagger}$ represents conjugate transpose,
$\vec{\theta} \rightarrow \{S^a_{ij}, S^p_{ij}, \Omega_{\alpha}, \alpha, \Delta \}$ represents our model parameters,
and the $\vec{\theta}$-dependant 3 $\times$ 3 covariance matrix $C(f_{k})$ is
\beq
C(f_{k}) =  \frac{\rm T_{tot}}{4}\left(\begin{array}{ccc}
\langle AA^{*} \rangle & \langle AE^{*} \rangle & \langle AT^{*} \rangle
\\
\langle EA^{*} \rangle & \langle EE^{*} \rangle & \langle ET^{*} \rangle
\\
\langle TA^{*} \rangle & \langle TE^{*} \rangle & \langle TT^{*} \rangle
\end{array} \right),
\eeq
includes contributions from both the instrumental noise spectral densities and signal spectral densities.

\subsection{Data simulation }

In our simulations, we focus on a simplified scenario and work under the following assumptions:
\begin{itemize}
\item[(1)] Our data are the sum of a SGWB and instrumental noise,
  i.e., we are working with ideal case, that all resolvable sources, glitches, and any other disturbances have been subtracted perfectly from the data.

\item[(2)] The noise and the signal are Gaussian, stationary, and uncorrelated in frequency domain.
Twelve noise components in the cross spcetra
are described by a model whose parameter values are known to within about $\pm 20\%$ of its nominal values.

\end{itemize}
The injected values of the SGWB model parameters can be found in Sec. \ref{sec:SGWB},
and the injected instrumental noise parameters are listed in the second column of Table~\ref{tab:results}.

For the sake of convenience, we generate data directly in the frequency domain, since this allows us to ignore window effects and overlapping segments.
We generate a three-month data set at 5 second sampling interval, which is related by a maximum frequency of 0.1 Hz.
The correlated noise are generated by
\begin{equation}
    \label{simulate}
    \begin{aligned}
        n_{A}(f_{k}) =& \frac{\sqrt{\left\langle N_{AA^{*}}\right\rangle }}{2}(x_1 + {\rm i}\, y_1),
        \\
        n_E(f_{k}) =& c_{1}\,n_A(f_{k})+ c_{2}\,(x_2 + {\rm i}\, y_2),
        \\
        n_T(f_{k}) =& c_{3}\,n_A(f_{k})+ c_{4}\,n_E(f_{k})
        + c_{5}\,(x_3 + {\rm i} \,y_3 ).
    \end{aligned}
\end{equation}
Likewise, independent signals for each channel can therefore be generated by
\begin{equation}
    \begin{aligned}
        h_A(f_{k}) =& \frac{ \sqrt{S_{\rm h}(f_{k})\,R_{AA^*}(f_{k})}}{2}
        \big(x_4 + {\rm i} \,y_4\big),
        \\
        h_E(f_{k}) =& \frac{\sqrt{S_{\rm h}(f_{k})\,R_{EE^*}(f_{k})}}{2}
        \big(x_5 + {\rm i}\, y_5\big),
        \\
        h_T(f_{k}) =& \frac{\sqrt{S_{\rm h}(f_{k})\,R_{TT^*}(f_{k})}}{2}
        \big(x_6 + {\rm i} \,y_6\big).
    \end{aligned}
\end{equation}
Here, $x_{i}$ and $y_{i}$, $i=\{1,2,3,4,5,6\}$, are statistically-independent real Gaussian random variables,
each of zero mean and unit variance, and the coefficients $\{c_{1}, c_{2}, c_{3}, c_{4}\}$ are derived in the Appendix \ref{Simulate}.

\begin{figure*}[!htpb]
    \centering
    \includegraphics[width=\textwidth]{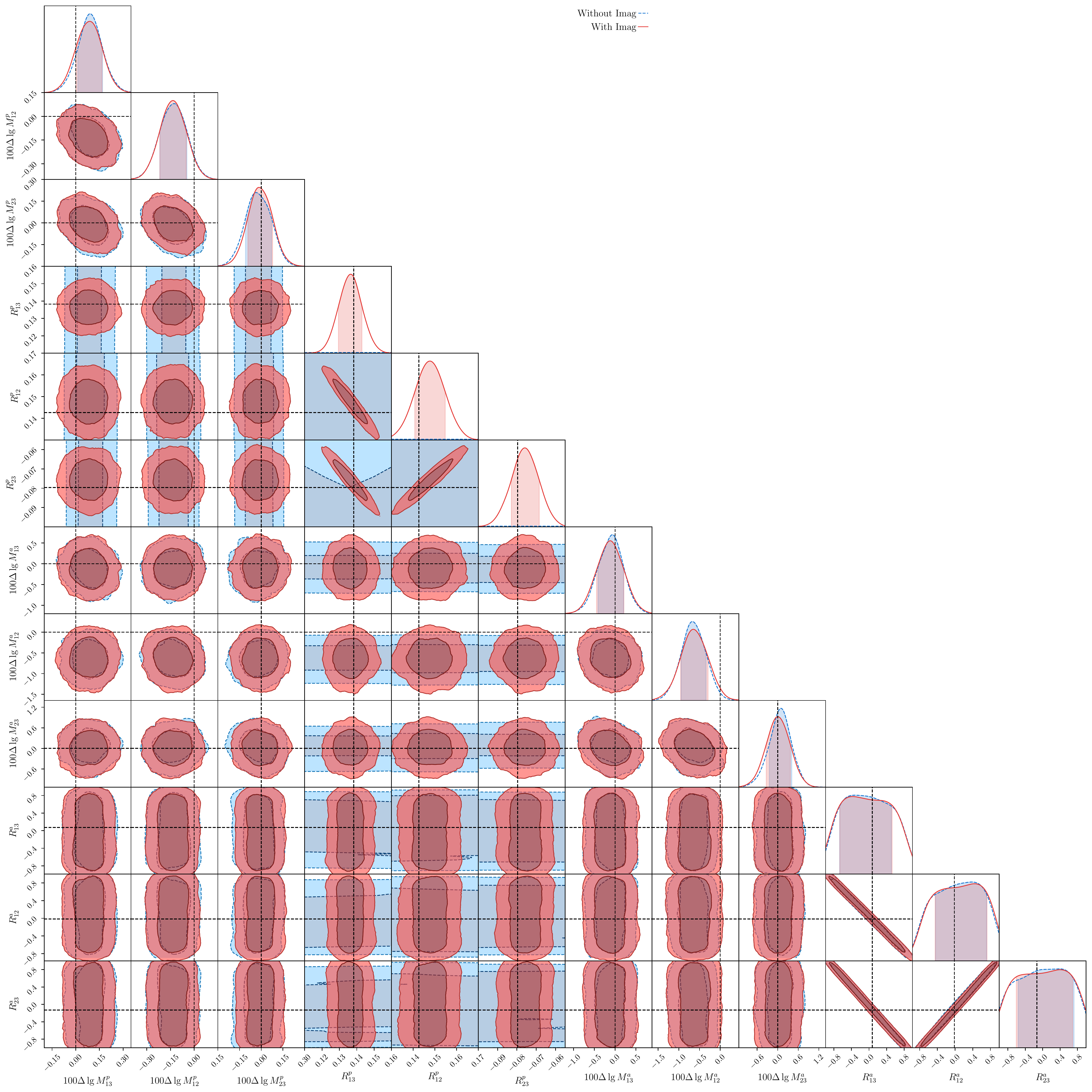}
    \caption{Corner plots of the noise-only data for the single TianQin configuration.
    The red (blue) contours and solid (dashed) lines are for the model with (without) imaginary components.
    The results are for the twelve TianQin noise parameters.
    The vertical dashed lines represent the injected values of the sum and ration of six position noises and six acceleration parameters,
    while the vertical shaded region on the posterior distribution denote 68\% credible region,
    the contour lines denote [68\% , 95\%] credible regions. }
\label{fig:Onlynoise_Complex}
\end{figure*}

\subsection{Parameter estimation}

The formalism of Eq.~\eqref{eq:lnlikeli} means that we can assess the noise parameters from the data.
In this subsection, we use \texttt{TQSGWB} to perform Bayesian inference on the \ac{GW} data under various models, and discuss the ability of assessing the position noise $S^{p}_{ij}$ and acceleration noise $S^{a}_{ij}$.
We first investigate the scenario where the data is composed with noise only.

From Eqs.~\eqref{tqspsa}, for the instrument noises, one has
\begin{equation}
    \begin{aligned}
        S^{a,p}_{ij} + S^{a,p}_{ji} =& \left( N^{a,p}_{ij} + N^{a,p}_{ji} \right) \\
        & \times 
    \begin{cases}
        \frac{\rm m^2}{\rm Hz} \left( \frac{1}{2L} \right)^2 & \text{for p}, \\
        \frac{\rm m^2}{\rm Hz} \left( 1 + \frac{0.1 \rm mHz}{f} \right) 
        \left( \frac{1}{2\pi f} \right)^4 \left( \frac{1}{2L} \right)^2 & \text{for a},
    \end{cases}
    \end{aligned}
\end{equation}
where the index $a,p$ represent acceleration and position noise, and $N^{a,p}_{ij}$ are the noise amplitude parameters of different arms.
Then we can define several new parameters with the sum and difference of noise parameters for each arm, i.e.,
\begin{align}
    M^{a,p}_{ij} =& N^{a,p}_{ij} + N^{a,p}_{ji}, \\
    D^{a,p}_{ij} =& N^{a,p}_{ij} - N^{a,p}_{ji},
\end{align}
and the ratio between the differences and the summations
\begin{align}
    R^{a,p}_{ij} =& \frac{D^{a,p}_{ij}}{M^{a,p}_{ij}} 
    = \frac{N^{a,p}_{ij} - N^{a,p}_{ji}}{N^{a,p}_{ij} + N^{a,p}_{ji}}.
\end{align}

\subsubsection{Constraint results on the instrument noise only}

In this scenario, we compare parameter estimation results under two different covariance matrix: with and without the imaginary components.
The contour plots of posterior for the parameters of $\Delta \lg M^{a,p}_{ij} = \lg M^{a,p}_{ij} - \lg M^{a,p}_{ij, \rm injected}$ and $R^{a,p}_{ij}$ are shown in Fig.~\ref{fig:Onlynoise_Complex}%
\footnote{The contour plots are generated using the plotting utilities \texttt{ChainConsumer} \cite{Hinton2016}.}.
Here, $\lg M^{a,p}_{ij, \rm injected}$ is the value derived from the injected values, i.e., see Table~\ref{tab:results}.
The injected values with dashed lines are also shown in the figures, and also the 68\% and 95\% credible regions with contour lines.
One should note that the simulated date are generated under the covariance with imaginary part, but two different models (with and without) are considered.
The contours in blue and dashed lines are for the case without imaginary components, the red contours and solid lines is for the case with imaginary part.
And the constraint results of different cases are listed in Table~\ref{tab:results}.

\begin{table}[htbp]
    \centering
    \caption{The injected values and constraint results for noise parameters.}
    \label{tab:results}
    \begin{tabular}{c|c|c|c}
        \hline
        \hline
        Parameter & Injected &  Without image & With image \\ 
        \hline
        $100 \Delta \lg M^p_{13}$ & 0 & $0.091^{+0.079}_{-0.078}$ & $0.089^{+0.078}_{-0.088}$ \\
        $100 \Delta \lg M^p_{12}$ & 0 & $-0.127^{+0.082}_{-0.090}$ & $-0.134^{+0.084}_{-0.084}$ \\
        $100 \Delta \lg M^p_{23}$ & 0 & $-0.035^{+0.107}_{-0.074}$ & $-0.016^{+0.094}_{-0.076}$ \\
        $R^p_{13}$ &   0.1382 & -- & $0.1365^{+0.0065}_{-0.0070}$ \\
        $R^p_{12}$ &   0.1426 & -- & $0.1478^{+0.007}_{-0.007}$ \\
        $R^p_{23}$ &  -0.0796 & -- & $ -0.0758^{+0.0074}_{-0.0070}$ \\
        $100 \Delta \lg M^a_{13}$ & 0 & $-0.06^{+0.27}_{-0.35}$ & $-0.12^{+0.33}_{-0.33}$ \\
        $100 \Delta \lg M^a_{12}$ & 0 & $-0.67^{+0.33}_{-0.27}$ & $-0.65^{+0.35}_{-0.31}$ \\
        $100 \Delta \lg M^a_{23}$ & 0 & $0.10^{+0.31}_{-0.34}$ & $0.00^{+0.37}_{-0.34}$ \\
        $R^a_{13}$ &  0.0735 & $-0.32^{+0.83}_{-0.35}$ & $-0.37^{+0.90}_{-0.30}$ \\
        $R^a_{12}$ & -0.0160 & $0.38^{+0.34}_{-0.82}$ & $0.44^{+0.28}_{-0.89}$ \\
        $R^a_{23}$ & -0.1318 & $0.38^{+0.35}_{-0.94}$ & $0.40^{+0.30}_{-1.00}$ \\
        \hline
        $\lg N^p_{13}$ & -23.9272 & $-23.85^{+0.12}_{-0.44}$ & $-23.93^{+0.0025}_{-0.0029}$ \\
        $\lg N^p_{12}$ & -23.9439 & $-23.89^{+0.15}_{-0.38}$ & $-23.94^{+0.0028}_{-0.0027}$ \\
        $\lg N^p_{23}$ & -24.0485 & $-23.88^{+0.10}_{-0.43}$ & $-24.05^{+0.0036}_{-0.0035}$ \\
        $\lg N^p_{31}$ & -24.0481 & $-23.86^{+0.11}_{-0.50}$ & $-24.05^{+0.0037}_{-0.0034}$ \\
        $\lg N^p_{21}$ & -24.0687 & $-23.91^{+0.16}_{-0.40}$ & $-24.07^{+0.0036}_{-0.0038} $ \\
        $\lg N^p_{32}$ & -23.9792 & $-23.89^{+0.13}_{-0.46}$ & $-23.98^{+0.0031}_{-0.0029} $ \\
        $\lg N^a_{13}$ & -29.9284 & $-29.85^{+0.16}_{-0.38}$ & $-29.86^{+0.16}_{-0.38}$ \\
        $\lg N^a_{12}$ & -29.9574 & $-29.81^{+0.14}_{-0.32}$ & $-29.80^{+0.12}_{-0.33}$ \\
        $\lg N^a_{23}$ & -30.0751 & $-29.90^{+0.15}_{-0.40}$ & $-29.89^{+0.13}_{-0.41}$ \\
        $\lg N^a_{31}$ & -29.9924 & $-29.83^{+0.15}_{-0.34}$ & $-29.82^{+0.13}_{-0.35}$ \\
        $\lg N^a_{21}$ & -29.9435 & $-29.88^{+0.17}_{-0.39}$ & $-29.87^{+0.16}_{-0.42}$ \\
        $\lg N^a_{32}$ & -29.9599 & $-29.90^{+0.15}_{-0.42}$ & $-29.91^{+0.14}_{-0.42}$ \\
        \hline
        \hline
    \end{tabular}
    \\
    \parbox[t]{0.92\linewidth}{\flushleft\footnotesize In the table, the symbols ``--'' represent that the parameters can not be constrained.}
\end{table}

\begin{figure}[htbp]
    \centering
    \includegraphics[width=0.8\linewidth]{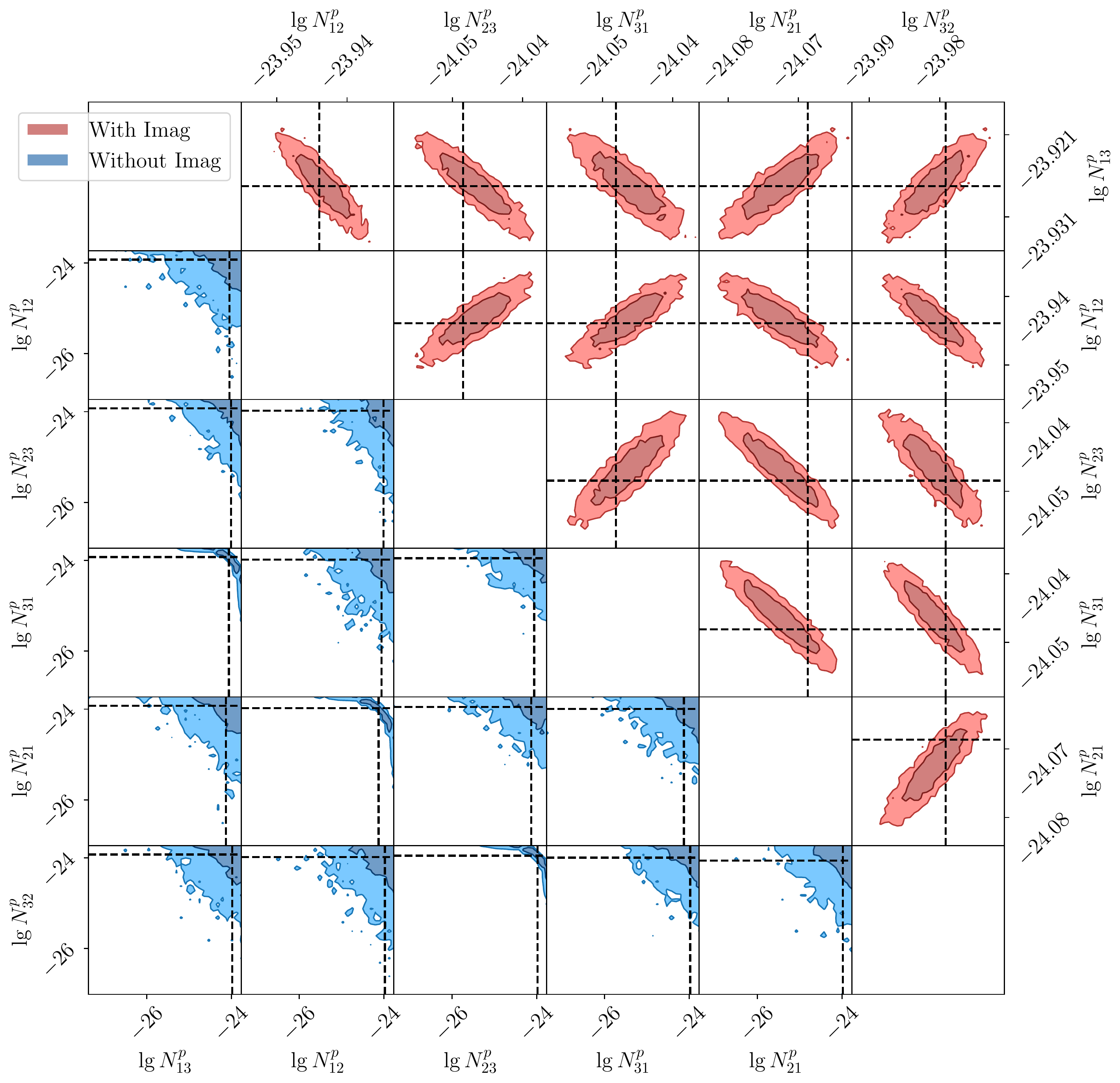}
    \caption{Contour plots of the six TianQin position noise parameters.
    The shaded area denote the 68\% and 95\% credible regions.
    The dashed line represent the injected values.}
    \label{fig:Onlynoise_pos}
\end{figure}

For the case of the model without imaginary part.
As expected, all summation parameters $M^{a,p}_{ij}$ are well-constrained, with slight negative correlation between different summations.
The latter can be explained by the expression in Appendix~\ref{Cross-Spectra}, which shows that for the diagonal terms (for both the position noises and the acceleration noises) all $M^{a,p}_{ij}$ adds up, leading to a negative correlation in between.
On the other hand, the ratio parameters $R^{a,p}_{ij}$ are much less constrained, which indicates that the differences parameters $D^{a,p}_{ij}$ are also less constrained.
Combining Figs.~\ref{fig:Onlynoise_Complex}, \ref{fig:Onlynoise_pos} and Table~\ref{tab:results}, one can find that though the constraint results for all $M^{a,p}_{ij}$ are better, except for the single parameters of $N^{a,p}_{ij}$, the credible regions are much larger than $M^{a,p}_{ij}$.
As shown in Fig.~\ref{fig:Onlynoise_pos}, there are big biases for the single noise parameters $N^p_{ij}$, which indicates that, in such case, it is only possible to constrain a certain combination of the noise parameters.

Once the imaginary components are included in the calculation, the degeneracies in position noises can be broken.
In Fig. \ref{fig:Onlynoise_pos}, we present the position noise parameters $N^{p}_{ij}$ \emph{per se}.
The panels in the upper-right triangle and lower-left triangle are for the contour plots of model with and without imaginary combinations, respectively.
As expected, each pair of position noises are negatively correlated, while correlation exist for other combinations.
And for the model without imaginary part, it is straightforward to see that only the contours of paired position noise parameters can be constrained.
However, for the model with imaginary part, all the position noise parameters are constrained tightly.

From Fig.~\ref{fig:Onlynoise_Complex} and Table~\ref{tab:results}, we observe that the 68\% credible region for the summation $M^{a,p}_{ij}$ is similar between models with or without imaginary components.
For the ratio parameters, covariance with imaginary components can give a good constraint on $R^p_{ij}$.
While the pattern of the contours remain roughly the same for $R^a_{ij}$, which is due to that there is no imaginary correlation contribute to the acceleration noise parameters.
From Table~\ref{tab:results} and Fig.~\ref{fig:Onlynoise_pos}, one can find that the 68\% credible region for the summation $\lg N^{a}_{ij}$ is larger than 0.1 without the imaginary correction, but shrinks to 0.003 with the imaginary correction.
This indicates that the credible region for the summation parameters are reduced by two orders of magnitude when the imaginary part is considered.

\subsubsection{Constraint results for different SGWB models}

\begin{figure}[bp]
    \centering
    \includegraphics[width=0.6\linewidth]{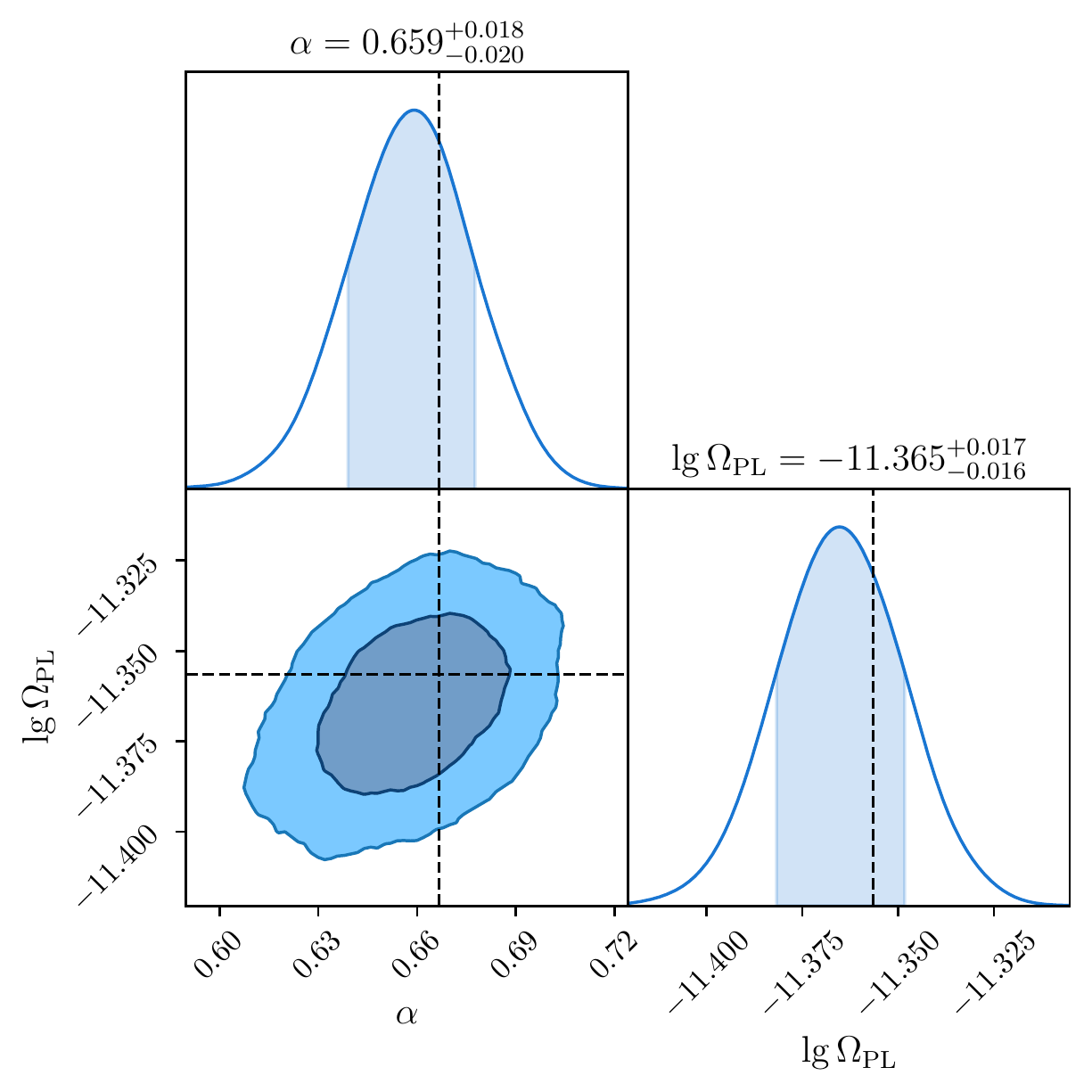}
    \caption{Corner plots for the noise + power-law SGWB case.
    Here only the parameters of the SGWB model are shown.
    The dashed lines represent the injected values.}
    \label{fig:PL_Omg}
\end{figure}

\begin{figure}[htbp]
    \centering
    \includegraphics[width=0.35\linewidth]{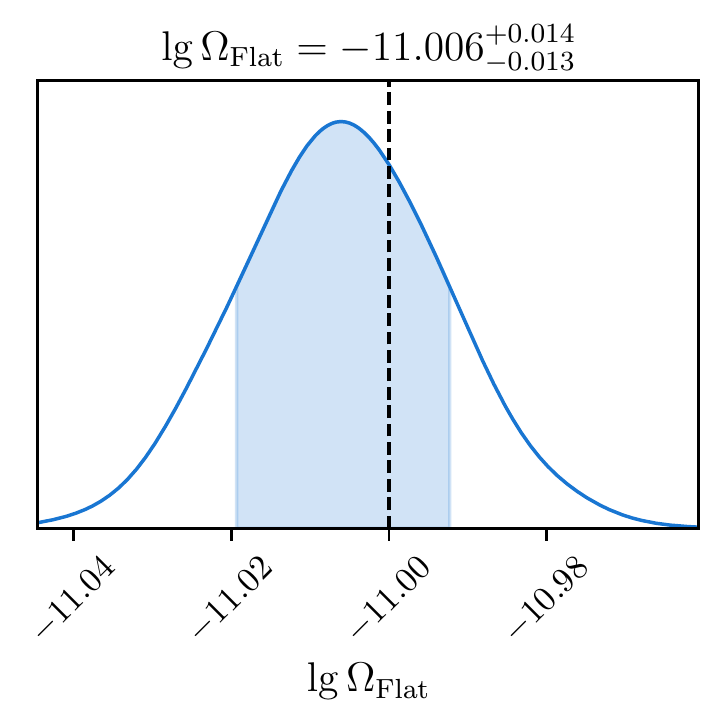}
    \caption{Posterior distribution of $\lg \Omega_{\rm Flat}$ for the noise + flat SGWB case.
    The vertical dashed line represents the injected value.}
    \label{fig:Flat_Omg}
\end{figure}

\begin{figure}[htbp]
    \centering
    \includegraphics[width=0.6\linewidth]{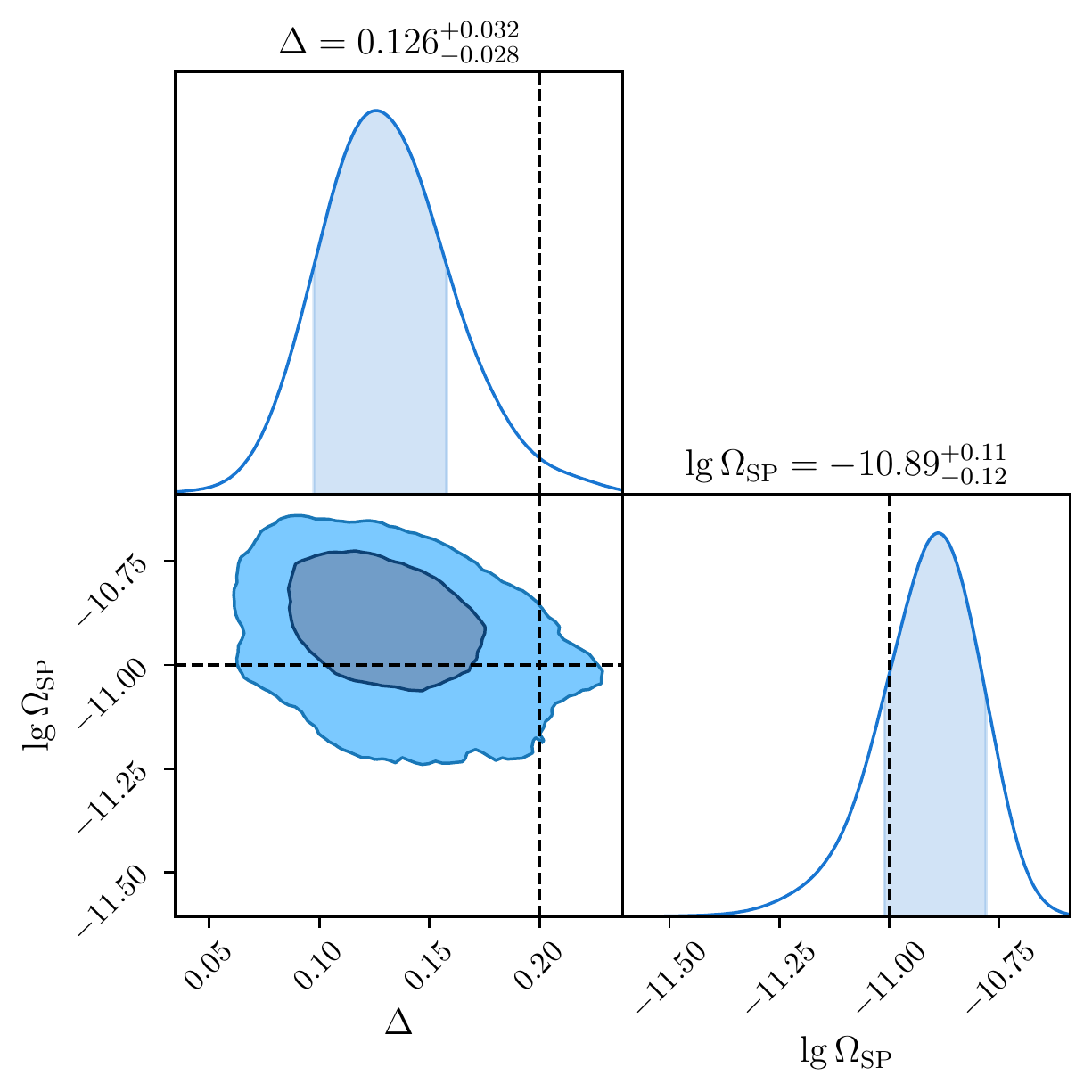}
    \caption{Corner plots for the noise + SP SGWB case.
    Here only the parameters of the SGWB model are shown.
    The dashed lines represent the injected values.}
    \label{fig:SP_Omg}
\end{figure}

We then turn to the scenarios when \ac{SGWB} is present in the data.
Since $A$, $E$, and $T$ channels have uncorrelated responses in the equal arm-length limit, 
diagonal elements in the covariance matrix, which contain signal and noise components, are used to distinguish \ac{SGWB} from instrument noise,
while off-diagonal terms are used to increase the constraint on the acceleration noise and position noise.
Similar to the noise-only scenario, by adopting the correct model, the injected noise parameters can be recovered precisely.

The constraint results for the parameters of the three SGWB models are shown in Figs.~\ref{fig:PL_Omg}, \ref{fig:Flat_Omg}, and \ref{fig:SP_Omg}, respectively.
Furthermore, the \ac{SGWB} parameters are also correctly recovered.
This results show the encouraging potential of our method to recover noise parameters like $N^{p}_{ij}$ and $N^{a}_{ij}$, as well as the \ac{SGWB} parameters like $\alpha$, and $\Omega$.
For the power-law SGWB model, the constraint precision of the amplitude parameter can reach the level of $\sim 0.2\%$, while for the index it can reach $\sim 3\%$.
For the flat SGWB model, the constraint precision of the amplitude parameter can approach $\sim 0.1\%$.
However, for the Gaussian-bump case, the injected value of the parameter $\Delta$ lies slightly outside the 68\% credible interval.
The constraint precisions for $\Omega_{\rm SP}$ and $\Delta$ reach the level of $\sim 1\%$ and $\sim 25\%$, respectively.
This can be partially explained by the low \ac{SNR} for the injected \ac{SGWB}.

We present contour plots between the SGWB model parameters and the instrumental noise parameters for a power-law, a flat, and a Gaussian-bump \ac{SGWB}, respectively in Figs. \labelcref{fig:PLnoise,fig:Flatnoise,fig:SPnoise}.
From these contour plots, one can find that, though the number of  model parameters have been increased, the precision and the biases of the constraint results for the instrument noises are basically unchanged.

\subsection{Detection limit}


We then shift attention to the Bayesian model selection.
In this part, we aim to answer to which level can we confidently use \texttt{TQSGWB} to identify the existence of \ac{SGWB}.
We consider two models:
\begin{itemize}
    \item[i.] $\mathcal{M}_{0}$: data is described by instrument noise only,
    \item[ii.] $\mathcal{M}_{1}$: data consists of instrument noise and an SGWB.
\end{itemize}
To compare the two models, we use the odds ratio, which numerically equals to the Bayes factor when we set the prior odds to be unit
\beq
{\cal B}^{1}_{0} = \frac{\int d{\vec{\theta}}~ {\cal L}(\vec{\theta}|D,{\cal M}_1 ) ~ p(\vec{\theta}\,|{\cal M}_1)}
{\int d{\vec{\theta}}~ {\cal L}(\vec{\theta}|D,{\cal M}_0 ) ~ p(\vec{\theta}\,|{\cal M}_0) },
\eeq
where $p(\vec{\theta}\,| {\cal M} ) $ and the integrand ${\cal L}(\vec{\theta}|D,{\cal M} )$ respectively represent the prior probability
and likelihood under the corresponding model.

A positive log Bayes factor $\log {\cal B}^{1}_{0}$ shows support for the ${\cal M}_1$ over ${\cal M}_0$.
But to avoid the influence from random fluctuation, it is widely suggested that a value of $\log {\cal B}^{1}_{0}>1$  is required for a meaningful followup discussion, and a value of $\log {\cal B}^{1}_{0}>3$ is needed for a strong support of the ${\cal M}_1$ \cite{Kass:1995loi}.

\begin{figure*}[!ht]
    \includegraphics[width=0.325\linewidth,trim={0.4cm 0 1.2cm 1.1cm}, clip]%
    {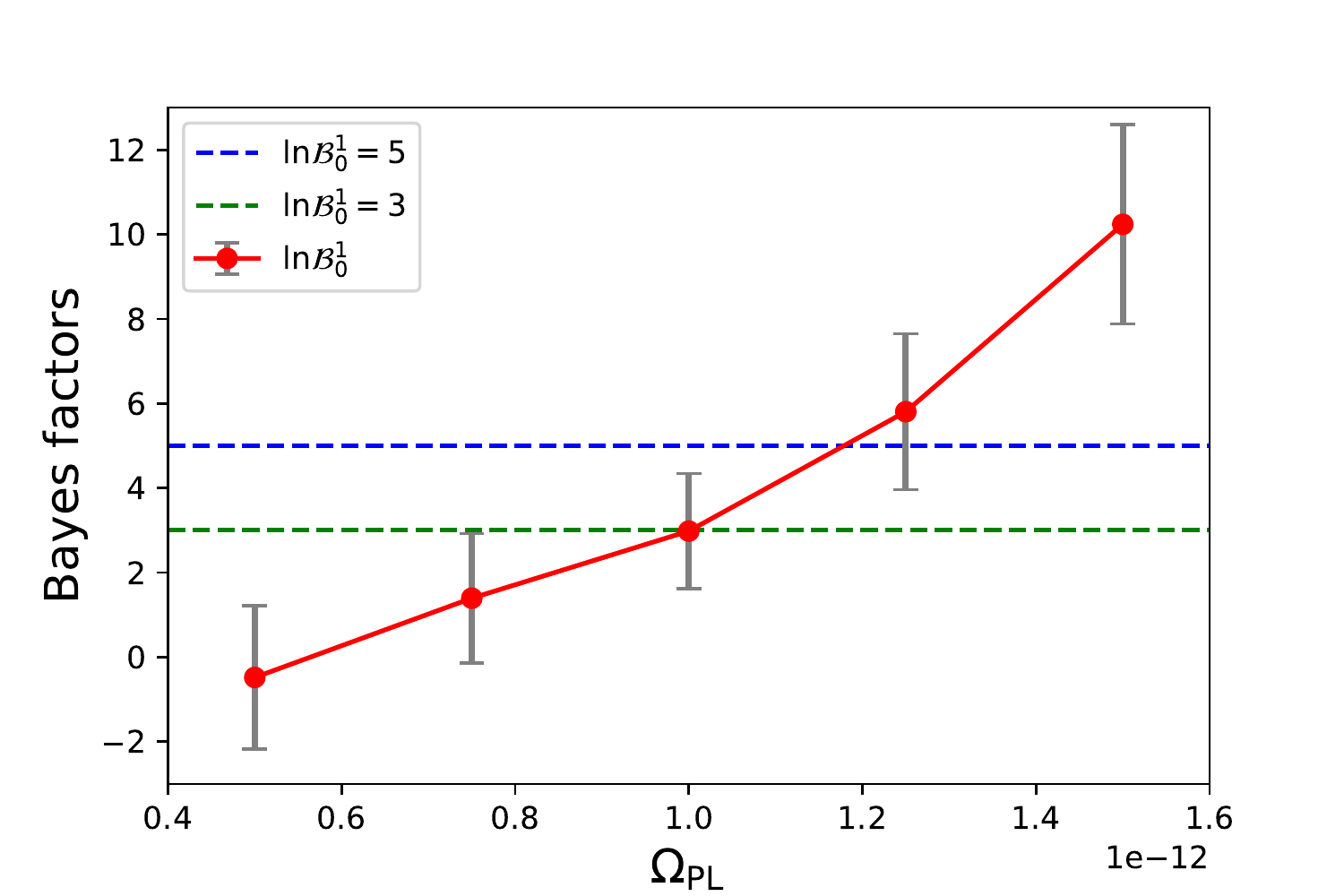}
    \includegraphics[width=0.325\linewidth,trim={0.4cm 0 1.2cm 1.1cm}, clip]%
    {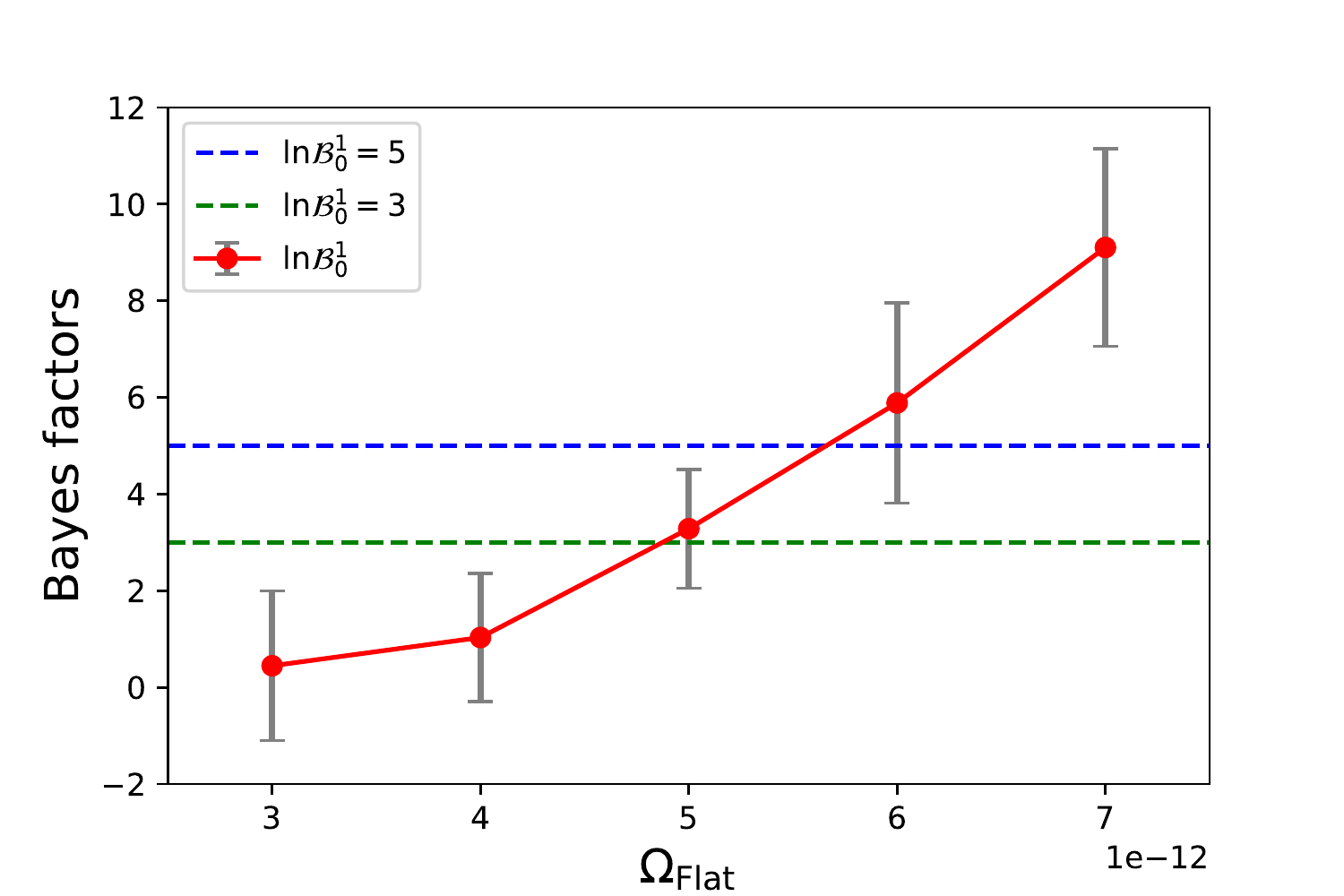}
    \includegraphics[width=0.325\linewidth,trim={0.4cm 0 1.2cm 1.1cm}, clip]%
    {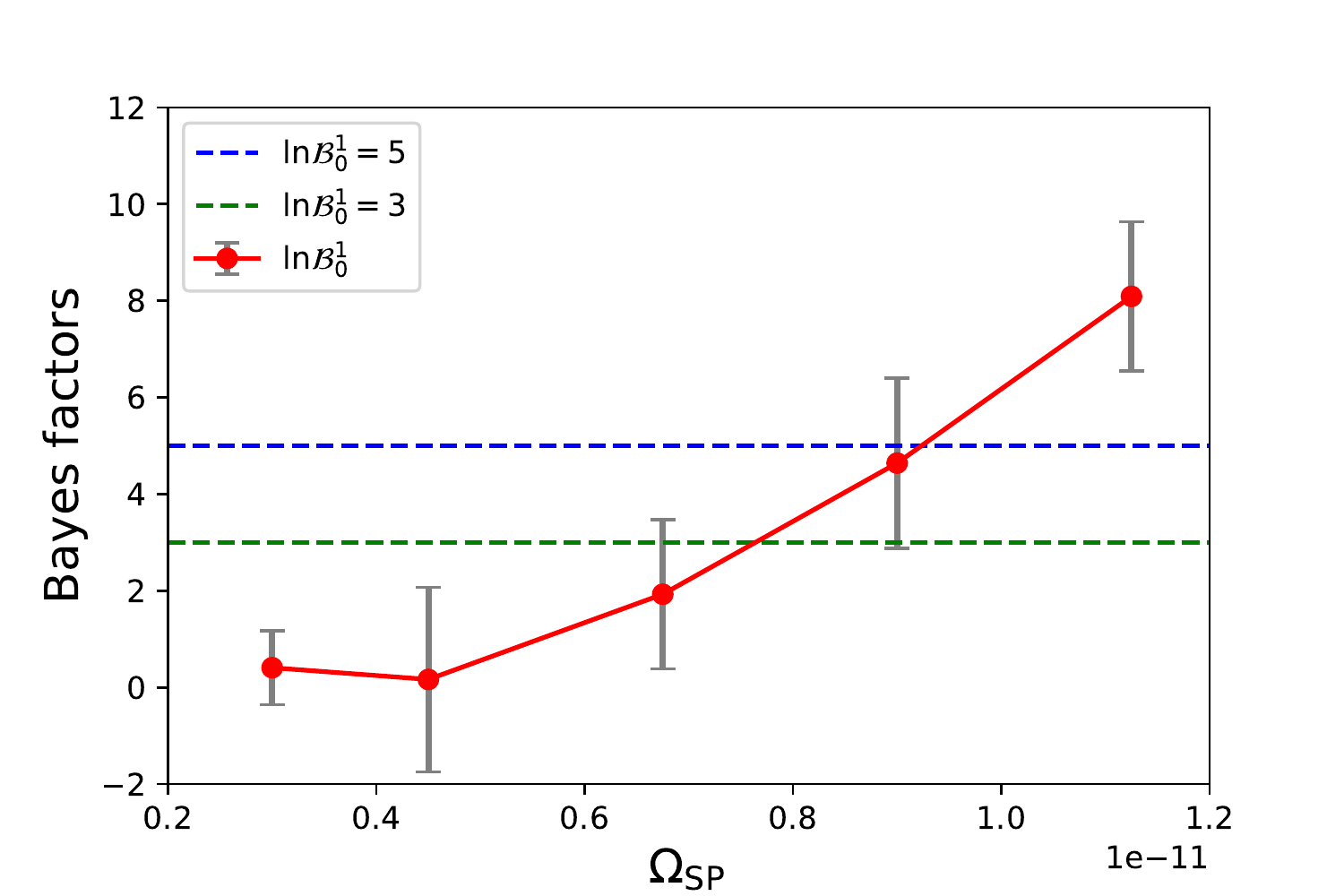}
  \caption{The Bayes factor as a function of background amplitude $\Omega_{0} $ in $A$, $E$, and $T$ channel,
which showing the detectability versus level of a flat background (left plane), a power law background (center plane)
and a single-peak background (right plane).
The green and blue dotted lines represent the log-Bayes factors 1 and 3, respectively.
  The magenta line represents the Bayes factor.}
\label{fig:BayesFactor}
\end{figure*}

In Fig. \ref{fig:BayesFactor}, we present log Bayes factor between the \ac{SGWB} models and the noise model, with the corresponding \ac{SGWB} injected in the data, and a total observation period of three months is assumed.
In each panel there are multiple horizontal lines, the green dashed line represents a log Bayes factor of 3, and the blue dashed line represents a log Bayes factor of 5.

Our detection confidence becomes very strong (a log Bayes factor of 5) for a power-law background level $\Omega_{\rm PL} = 1.3 \times 10^{-12}$, a flat background level of $\Omega_{\rm Flat} = 6.0 \times 10^{-12}$,
and a single-peak background level $\Omega_{\rm SP} = 9.0 \times 10^{-12}$
with three-month data.
The corresponding \acp{SNR} are around 5, 13, and 8, respectively.

\section{SUMMARY AND DISCUSSION}\label{sec:SUMMARY AND DISCUSSION}

In this paper, we investigated the capability of TianQin to detect various isotropic \ac{SGWB} using the Bayes analysis method, where a full covariance matrix or cross-correlation between different detection channels is considered in the likelihood function.
To the best of the authors' knowledge, for the first time, we include the imaginary components in the cross spcetral \ac{TDI} channels covariance matrix, which raised due to the asymmetrical instrument noises.
We perform a Bayesian inference under a number of different cases to demonstrate the validity of the updated covariance matrix.
We also applied a model selection analysis to show how sensitive is TianQin to the strength of \ac{SGWB} model.
Three models are considered for the  \ac{SGWB} spectrum: the \acp{CBC} (power-law), a scale invariant inflationary \ac{SGWB} (flat), and a single-peak spectrum (Gaussian-bump).
Throughout our analysis, we have assumed the \ac{SGWB} and noise to be Gaussian and stationary, and we assume an operation time of three months due to TianQin's (3+3)-operational model.

We have calculated the cross spcetral density of instrumental noise working in the TDI-1.5 channels $A,$ $E,$ and $T$.
Applying the MCMC on noise-only data, we demonstrated that including the imaginary components can help to break degeneracies between different noise parameters, especially for the position noise.
Similar conclusions are also observed when three representative \ac{SGWB} are injected, 
which demonstrated that both the noise and \ac{SGWB} parameters can be recovered precisely under the our model.

To assess the detection limit of TianQin to \ac{SGWB}, we apply Bayesian model selection to noise and \ac{SGWB} model versus the noise only model.
We compute the Bayes factors under different \ac{SGWB} strengths, and concluded that TianQin will be able to detect the energy density of
a PL signal as low as $\Omega_{\rm PL} = 1.3 \times 10^{-12}$,
a flat signal as low as $\Omega_{\rm Flat} = 6.0 \times 10^{-12}$,
and a SP signal as low as $\Omega_{\rm SP} = 9.0 \times 10^{-12}$ with three-month observation.

The primary purpose of this paper is to explore the detection capability of the SGWB via TianQin, and focusing on the impacts of including the imaginary component of the covariance matrix.
What is more, the discussion here focuses on distinguishing single-component SGWB from instrument noise,
and does not yet address issues of identifying astrophysical and cosmological sources simultaneously.

One line of our future study is to extend to multiple independent space-based GW detectors, such as TianQin-LISA network.
TianQin and LISA are two mHz band space-based GW detectors which will probably fly at roughly the same time, around 2035.
With this network, one can expect to apply the cross-correlation method to detect \ac{SGWB}.
Another line is to see how the limit is affected by astrophysical confusion foregrounds, such as the noise from un-resolved white dwarf binaries in our Galaxy.
We leave those for future works.

\begin{acknowledgments}
The authors would like to thank Pan-Pan Wang for discussing response function and Zhiyuan Li for useful conversations.
    This work has been supported by the National Key Research and Development Program of China (Grant No. 2020YFC2201400),
    the Guangdong Major Project of Basic and Applied Basic Research (Grant No. 2019B030302001),
    the Natural Science Foundation of China (Grant No. 12173104),
    the fellowship of China Postdoctoral Science Foundation (Grant No. 2021M703769),
    and the Natural Science Foundation of Guangdong Province of China (Grant No. 2022A1515011862).
    We acknowledge the support by National Supercomputer Center in Guangzhou.
\end{acknowledgments}

\begin{widetext}

\appendix

\section{Cross spectra }\label{Cross-Spectra}

Here we provide a detailed calculation with respect to the noise cross spcetra between different TDI channels.

Due to the absence of correlation between different links as well as types of noise, the one-sided Gaussian spectral density of the noise can be defined as~\cite{Vallisneri:2012np}
\beq
\langle \widetilde{n}_{ij}^{\alpha}(f)\widetilde{n}_{kl}^{\beta*}(f')\rangle=\frac{1}{2}\delta_{ij,kl}\delta_{\alpha,\beta}\delta(f-f')S_{ij}^{\alpha}(f),
\eeq
where $\alpha$ and $\beta$ are employed to label the position
noise and the acceleration noise. Based on the construction of the $AET$ channel group (i.e., Eq.~(\ref{AET})), the noise cross spcetra are given by
\begin{align}
    \left\langle N_{AA^*}^p \right\rangle
    =& 2 \sin^2 u \bigg\{
    2 \big[1+ \cos u \big] \left( S_{13}^p+S_{31}^p \right)
    + \left( S_{12}^p+S_{21}^p \right)
    + \left( S_{23}^p+S_{32}^p \right) \bigg\}
    \nonumber \\
    \left\langle N_{AA^*}^a \right\rangle
    =& 8 \sin^2 u \bigg\{
    \big[ 1 + \cos u \big]^2 \left(S_{13}^a+S_{31}^a\right)
    + \cos^2 u \left( S_{21}^a+S_{23}^a \right)
    + \left(S_{12}^a+S_{32}^a\right) \bigg\}
    \nonumber \\
    \left\langle N_{EE^*}^p \right\rangle
    =& \frac{2}{3} \sin^2 u \bigg\{
    2 \big[ 1- \cos u \big] (S_{13}^p + S_{31}^p )
    + \big[ 5 + 4\cos u \big] (S_{12}^p +S_{21}^p +S_{23}^p+S_{32}^p) \bigg\}
    \nonumber \\
    \left\langle N_{EE^*}^a \right\rangle
    =& \frac{8}{3} \sin^2 u \bigg\{
    \big[ 1 +2\cos^2 u \big]^2
        \left(S_{12}^a +S_{32}^a \right)
    + \big[2 + \cos^2 u ]^2
        \left(S_{21}^a+S_{23}^a \right)
    + \big[1 - \cos u]^2
        \left(S_{13}^a + S_{31}^a \right)\Big\}
    \nonumber \\
    \left\langle N_{TT^*}^p \right\rangle
    =& \frac{16}{3} \sin^2 u \sin ^2\left(\frac{u}{2}\right)
    \left(S_{12}^p+S_{21}^p+S_{13}^p+S_{31}^p+S_{23}^p+S_{32}^p\right)
    \nonumber \\
    \left\langle N_{TT^*}^a \right\rangle
    =& \frac{64}{3}  \sin^2 u \sin ^4 \left(\frac{u}{2}\right)
    \left(S_{12}^a+S_{21}^a+S_{13}^a+S_{31}^a+S_{23}^a+S_{32}^a\right)
    \nonumber \\
    \left\langle N_{AE^*}^p \right\rangle
    =& \frac{2}{\sqrt{3}}\sin^2 u \bigg\{
        \big[1 + 2\cos u \big]
        ( S_{23}^p+S_{32}^p- S_{12}^p-S_{21}^p )
    + {\color{red} {\rm i} 2 \sin u
    \left(S_{12}^p-S_{21}^p-S_{13}^p+S_{31}^p+S_{23}^p-S_{32}^p\right)}\bigg\}
    \nonumber \\
    \left\langle N_{AE^*}^a \right\rangle
    =& \frac{8 }{\sqrt{3}} \sin^2 u \bigg\{
    \sin^2 u \left(S_{31}^a-S_{13}^a \right)
    + \cos u \big[2 +\cos u \big] \left(S_{23}^a-S_{21}^a\right)
    + \big[ 1+2\cos u \big] \left( S_{32}^a - S_{12}^a \right)
    \bigg\}
    \nonumber \\
    \left\langle N_{AT^*}^p \right\rangle
    =& 2 \sqrt{\frac{2}{3}} \sin^2 u \bigg\{
        \big[ 1-\cos u \big] \left(S_{23}^p+S_{32}^p -S_{12}^p-S_{21}^p\right)
        - {\color{red} {\rm i} \sin u
        \left(S_{12}^p -S_{21}^p + S_{23}^p-S_{32}^p +2 S_{13}^p-2 S_{31}^p \right)}
    \bigg\}
    \nonumber \\
    \left\langle N_{AT^*}^a \right\rangle
    =& 8 \sqrt{\frac{2}{3}} \sin^2 u \bigg\{
        \sin^2 u \left(S_{13}^a -S_{31}^a \right)
        + \cos u\big[ \cos u -1 \big] \left( S_{23}^a - S_{21}^a\right)
        + 2\big[ \cos u -2 \big] \left( S_{12}^a - S_{32}^a \right)
    \bigg\}
    \nonumber \\
    \left\langle N_{ET^*}^p \right\rangle
    =& \frac{2 \sqrt{2}}{3} \sin^2 u \bigg\{
        \big[ 1-\cos u \big]
        \left( 2 S_{13}^p + 2 S_{31}^p - S_{21}^p -S_{12}^p - S_{23}^p - S_{32}^p \right)
        - {\color{red} {\rm i} 3 \sin u
        \left(S_{21}^p-S_{12}^p + S_{23}^p - S_{32}^p\right)}\bigg\}
    \nonumber \\
    \left\langle N_{ET^*}^a\right\rangle
    =& \frac{8 \sqrt{2}}{3} \sin^2 u \big[\cos u - 1\big] \bigg\{
        \big[\cos u -1 \big]
        \left( S_{13}^a+ S_{31}^a \right)
        + \big[ \cos u +2 \big]
        \left( S_{21}^a + S_{23}^a \right)
        + \big[2\cos u +1 \big]
        \left( S_{12}^a + S_{32}^a \right)
        \bigg\}
    \label{eq:Np_CS}
\end{align}	
According to the above equations, if we assume symmetry between instrumental noises, then we can safely ignore imaginary terms,
However, once we abandon the symmetric assumption, one can find that it is no longer safe to ignore the imaginary part.
Take $\langle N^p_{AT^*} \rangle$ and $\langle N^p_{ET^*} \rangle$ for example, the imaginary part is of the order $\sim u$, while the real part is of the order $(1-\cos u)$,
so in the low frequency limit, the imaginary terms have a higher impact than the real terms.

Assuming that the three arm-lengths are equal and the noise in each satellite is exactly identical,
the corresponding channel noise power spectral density can be analytically expressed as
\begin{align}\label{AAEETTpsd}
\langle N_{AA} \rangle =  \langle N_{EE} \rangle &= 8 \sin^2 u
\bigg\{\big[2+\cos u\big] S^{p}+4\big[1+\cos u+\cos^2 u\big] S^{a}\bigg\},
\nonumber \\
\langle N_{TT} \rangle &= 16 \sin^2 u\bigg\{\big[1-\cos u\big]S^{p}
+2\big[1-\cos u\big]^2 S^{a}\bigg\}.
\end{align}

\section{Data generation}\label{Simulate}

We derive in detail here the values of the coefficients that enter expressions (\ref{simulate}),
\beqa
\left\langle n_{A}(f_{k})\, n^{*}_E(f_{k})\right\rangle  &=& \left\langle n_{A}(f_{k}) n^{*}_A(f_{k}) \right\rangle  c^{*}_{1} \,
\equiv \frac{1}{2}\left\langle N_{AE^{*}}\right\rangle ,
\nonumber \\
\left\langle n_{E}(f_{k})\, n^{*}_E(f_{k})\right\rangle  &=& \left\langle n_{A}(f_{k}) n^{*}_A(f_{k}) \right\rangle  |c_{1}|^2 \,+ 2|c_{2}|^2\,
\equiv \frac{1}{2}\left\langle N_{EE^{*}}\right\rangle ,
\nonumber \\
& \Rightarrow &
c_1 = \frac{\left\langle N_{EA^{*}}\right\rangle }{\left\langle N_{AA^{*}}\right\rangle }, \,~~
c_2 = \frac{\sqrt{\left\langle N_{EE^{*}}\right\rangle  - \left\langle  N_{AA^{*}}\right\rangle  \, |c_1|^2}}{2} \, .
\eeqa
Analogously, the parameters for simulating $n_{T}(f_{k})$ are derived
\beqa
\left\langle n_{A}(f_{k})\, n^{*}_T(f_{k})\right\rangle  &=& \left\langle n_{A}(f_{k}) n^{*}_A(f_{k}) \right\rangle  c^{*}_{3}
\,+\left\langle n_{A}(f_{k})\, n^{*}_E(f_{k})\right\rangle  c^{*}_{4} \,
\equiv \frac{1}{2}\left\langle N_{AT^{*}}\right\rangle  ,
\nonumber \\
\left\langle n_{E}(f_{k})\, n^{*}_T(f_{k})\right\rangle  &=& \left\langle n_{E}(f_{k}) n^{*}_A(f_{k}) \right\rangle  c^{*}_{3}
\,+ \left\langle n_{E}(f_{k})\, n^{*}_E(f_{k})\right\rangle c^{*}_{4}\,
\equiv \frac{1}{2} \left\langle N_{ET^{*}}\right\rangle ,
\nonumber \\
\left\langle n_{T}(f_{k})\, n^{*}_T(f_{k})\right\rangle  &=& \left\langle n_{T}(f_{k}) n^{*}_A(f_{k}) \right\rangle  c^{*}_{3}
\,+\left\langle n_{T}(f_{k})\, n^{*}_E(f_{k})\right\rangle  c^{*}_{4} \, + 2|c_{5}|^2
\equiv \frac{1}{2}\left\langle N_{TT^{*}}\right\rangle ,
\nonumber \\
& \Rightarrow &
c_{3} = \sqrt{\frac{\left\langle N_{TA^{*}}\right\rangle  \left\langle N_{EE^{*}}\right\rangle - \left\langle N_{EA^{*}}\right\rangle  \left\langle N_{TE^{*}}\right\rangle }
{\left\langle N_{AA^{*}}\right\rangle  \left\langle N_{EE^{*}}\right\rangle - \left\langle N_{AE^{*}}\right\rangle \left\langle N_{EA^{*}}\right\rangle  }}\,,
\nonumber \\
&& c_{4}= \sqrt{\frac{\left\langle N_{AA^{*}}\right\rangle  \left\langle N_{TE^{*}}\right\rangle - \left\langle N_{TA^{*}}\right\rangle  \left\langle N_{AE^{*}}\right\rangle }
{\left\langle N_{AA^{*}}\right\rangle  \left\langle N_{EE^{*}}\right\rangle - \left\langle N_{AE^{*}}\right\rangle \left\langle N_{EA^{*}}\right\rangle  }}\,,
\nonumber \\
&& c_{5}= \frac{\sqrt{\left\langle N_{TT^{*}}\right\rangle -c_{3}\left\langle N_{AT^{*}}\right\rangle - c_{4}\left\langle N_{ET^{*}}\right\rangle  }}{2}.
\eeqa

\section{Parameter estimation with different SGWB models}\label{PE_SGWB}

Here we present the parameter estimation results.
We inject different types of \acp{SGWB}, and recover under the same \ac{SGWB} models.

\begin{figure*}[!htpb]
\includegraphics[width=1.0 \textwidth,angle=0]{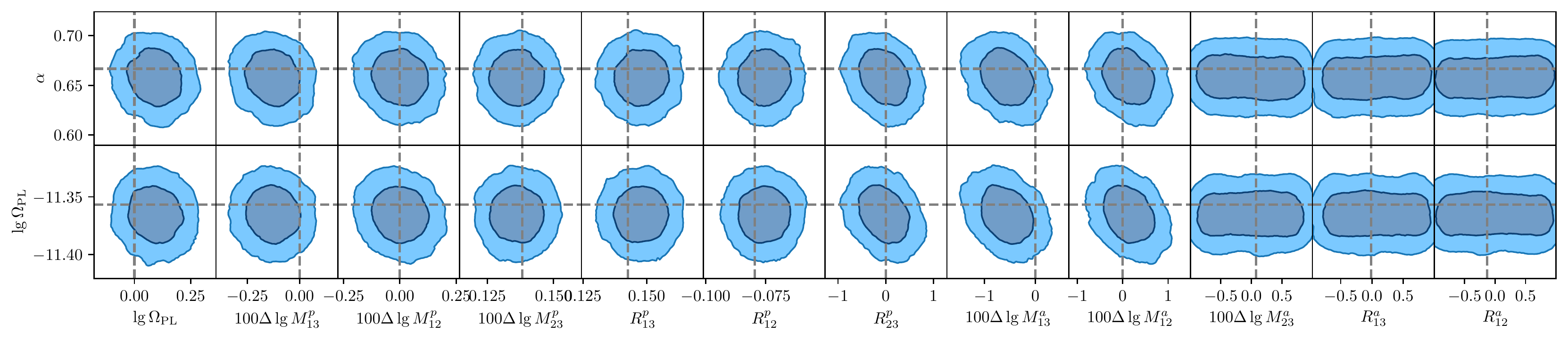}
\caption{Contour plot for the noise + power-law SGWB case for single TianQin configuration.
The vertical lines represent the injected values of the twelve noise parameters and two SGWB parameters (amplitude and spectral slope),
while the vertical dashed lines on the posterior distribution denote from left to right the quantiles [16\% , 84\%]. }
\label{fig:PLnoise}
\end{figure*}

\begin{figure*}[!htpb]
\includegraphics[width=1.0 \textwidth,angle=0]{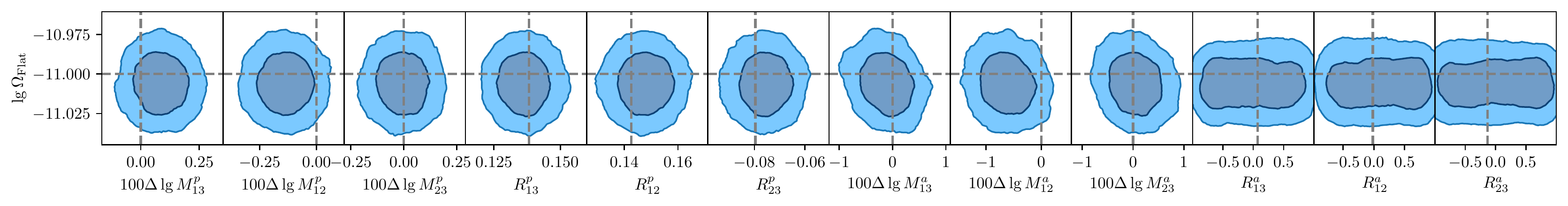}
\caption{Contour plot for the  noise+flat SGWB case for single TianQin configuration.
The vertical line represents the injected values of the 12 noise parameters,
while the vertical dashed lines on the posterior distribution denote from left to right the quantiles [16\% , 84\%]. }
\label{fig:Flatnoise}
\end{figure*}

\begin{figure*}[!htpb]
\includegraphics[width=1.0 \textwidth,angle=0]{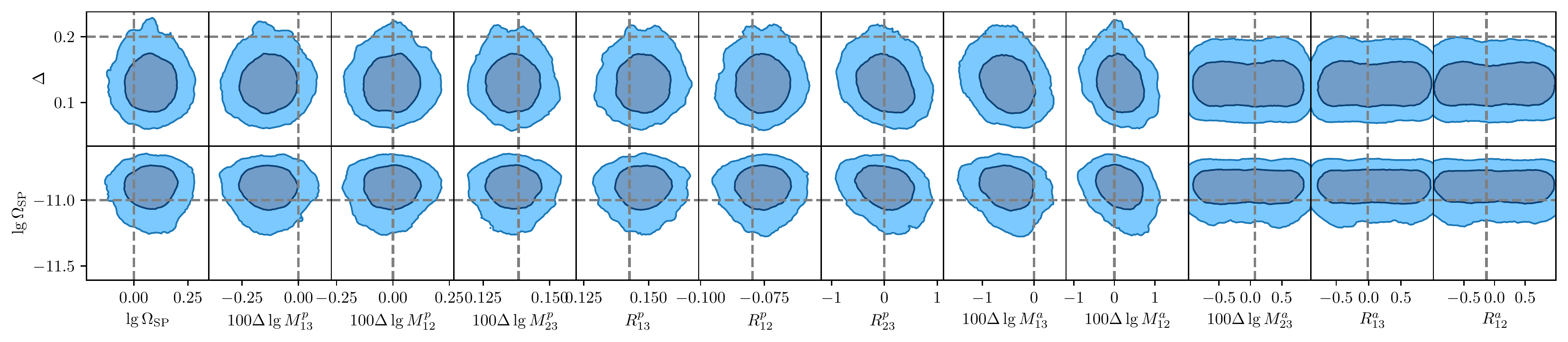}
\caption{Contour plot for the noise+SP SGWB case with single TianQin configuration.
The vertical line represents the injected values of the 12 noise parameters,
while the vertical dashed lines on the posterior distribution denote from left to right the quantiles [16\% , 84\%]. }
\label{fig:SPnoise}
\end{figure*}

\end{widetext}

\bibliography{SGWB}

\end{document}